\documentclass{test}

\makeatletter
\def\mdseries@tt{m}
\makeatother

\usepackage[frozencache=true, cachedir=.]{minted}

\usepackage{graphicx}
\usepackage{textcomp}
\usepackage{xcolor}
\usepackage{subfigure}
\usepackage{todonotes}
\usepackage{url}
\usepackage{lineno,hyperref}
\usepackage{amsmath}
\usepackage{amssymb}
\usepackage[numbers]{natbib}
\usepackage{cleveref}
\usepackage{soul}

\newcommand{\textsourcecode}[1]{{\fontsize{9}{9}\selectfont\sffamily\upshape #1}}

\usepackage{todonotes}

\usepackage{listings}
\usepackage{multirow}

\definecolor{my_white}{rgb}{1, 1, 1}                
\definecolor{my_black}{rgb}{0, 0, 0}                
\definecolor{my_blue}{rgb}{0, 0.125, 0.376}         
\definecolor{my_green}{rgb}{0.12, 0.3, 0.17}        
\definecolor{my_violet}{rgb}{0.44, 0.16, 0.39}      

\lstdefinestyle{PyCOMPSs}{
  backgroundcolor=\color{my_white},   
  basicstyle=\footnotesize,        
  breakatwhitespace=false,         
  breaklines=true,                 
  captionpos=None,                    
  commentstyle=\color{my_green},   
  frame=single,	                   
  keepspaces=true,                 
  keywordstyle=\color{my_blue},    
  language=Python,                 
  numbers=none,                    
  numbersep=5pt,                   
  numberstyle=\tiny\color{my_black},   
  rulecolor=\color{my_black},      
  showspaces=false,                
  showstringspaces=false,          
  showtabs=false,                  
  stepnumber=1,                    
  stringstyle=\color{my_violet},   
  tabsize=2,	                   
  morekeywords={@task, @constraint, compss_wait_on},             
}

\usepackage{fancyhdr}
\begin{document}

\shorttitle{Towards Enabling I/O Awareness in Task-based Programming Models}
\shortauthors{Elshazly et~al.}

\title [mode = title]{Towards Enabling I/O Awareness in Task-based Programming Models}                      
\tnotemark[1,2]

\tnotetext[1]{DOI: \url{10.1016/j.future.2021.03.009}}

\tnotetext[2]{\copyright <2021> Elsevier. This manuscript version is made available under the CC-BY-NC-ND 4.0 license \url{http://creativecommons.org/licenses/by-nc-nd/4.0/}
}

\author[1]{Hatem Elshazly}[orcid=0000-0002-8591-1502]
\cormark[1]
\ead{hatem.elshazly@bsc.es}

\address[1]{Barcelona Supercomputing Center (BSC), Barcelona, Spain.}

\author[1]{Jorge Ejarque}[orcid=0000-0003-4725-5097]
\ead{jorge.ejarque@bsc.es}

\author[1]{Francesc Lordan}[orcid=0000-0002-9845-8890]
\ead{francesc.lordan@bsc.es}

\author[1]{Rosa M. Badia}[orcid=0000-0003-2941-5499]
\ead{rosa.m.badia@bsc.es}

\cortext[cor1]{Corresponding author}

\begin{abstract}
   Storage systems have not kept the same technology improvement rate as computing systems. As applications produce more and more data, I/O becomes the limiting factor for increasing application performance. I/O congestion caused by concurrent access to storage devices is one of the main obstacles that cause I/O performance degradation and, consequently, total performance degradation. 
   
   Although task-based programming models made it possible to achieve higher levels of parallelism by enabling the execution of tasks in large-scale distributed platforms, this parallelism only benefited the compute workload of the application. Previous efforts addressing I/O performance bottlenecks either focused on optimizing fine-grained I/O access patterns using I/O libraries or avoiding system-wide I/O congestion by minimizing interference between multiple applications. 
   
   In this paper, we propose enabling \textit{I/O Awareness} in task-based programming models for improving the total performance of applications. An I/O aware programming model is able to create more parallelism and mitigate the causes of I/O performance degradation. On the one hand, more parallelism can be created by supporting special tasks for executing I/O workloads, called \textit{I/O tasks}, that can overlap with the execution of compute tasks. On the other hand, I/O congestion can be mitigated by constraining I/O tasks scheduling. We propose two approaches for specifying such constraints: explicitly set by the users or automatically inferred and tuned during application's execution to optimize the execution of variable I/O workloads on a certain storage infrastructure.
   
   We implement our proposal using PyCOMPSs: a Task-based programming model for parallelizing Python applications. Our experiments on the MareNostrum 4 Supercomputer demonstrate that using I/O aware PyCOMPSs can achieve significant performance improvement in the total execution time of applications with different I/O workloads. This performance improvement can reach up to 43\% of total application performance as compared to the I/O non-aware version of PyCOMPSs. 
\end{abstract}


\maketitle

\section{Introduction}
\label{sec:intro}

~\

The continuous growth in computing power has the ability to deliver increasing levels of parallelism to satisfy the computing demands of scientific applications. This increase in computing power not only has benefited from on-chip processors scaling but also from increasing the number of computing nodes in production systems. Today's supercomputers typically are compromised of hundreds of thousands of computing nodes enabling near exascale performance \cite{top500}. 

In order to harness this increasing computing power and turn it into performance improvements for applications, task-based programming models offer a flexible approach for parallelizing and executing applications in distributed platforms. Using a task-based execution approach, an application is decomposed into work units called \textit{Tasks}. Tasks are organized into an execution graph by detecting data dependencies between them. Later, they are executed if and only if all of their data dependencies are satisfied. Unlike programming models that follow a rigid parallel paradigm such as Map-Reduce \cite{dean2008mapreduce} or Spark \cite{spark}, task-based programming models allow for easier expression of the irregular and unstructured parallelism patterns of scientific applications. Other parallel programming models such as MPI \cite{gropp1999using} are widely used. However, gaining performance requires knowledge about the underlying execution infrastructure which could compromise applications programmability. 


Along with the demand for computational power, scientific applications also tend to be I/O intensive \cite{bigout}. Applications in critical areas such as computational biology and climate science generate large amounts of data usually for checkpointing intermediate results and restarting after failures \cite{checkpointing}, or for performing post-processing operations such as visualization and post-mortem analysis \cite{visualization}. The life cycle of these applications typically alternates between a computing phase followed by an I/O phase. During the I/O phase, large amounts of concurrent I/O requests overwhelm the I/O bandwidth of the storage system causing I/O congestion. I/O congestion was observed to cause significant slowdown in the I/O performance of applications \cite{iocongestion}. Indeed, I/O performance slowdown consequently degrades applications total performance.

Improvements in storage devices such as Burst Buffers \cite{burstbuffer} have been introduced to supercomputers to absorb the I/O of applications. However, as the amount of data generated by applications continues to grow, relying only on Burst Buffers is not enough to completely hide or mitigate I/O congestion. This performance gap between storage systems performance and the I/O requirements of applications has resulted in I/O becoming the bottleneck that prevents achieving more performance improvements for applications.


Current task-based programming models do not offer support targeting the I/O bottleneck of I/O intensive applications. Addressing the I/O bottleneck is conventionally done using low-level I/O libraries and middleware (e.g. MPI-IO \cite{mpiio}, HDF5~\cite{hdf5}). I/O libraries typically focus on parallelizing I/O access or optimizing I/O access patterns of applications. However, this approach is limited, as it cannot take advantage of coarse-grained performance improvements opportunities. Additionally, it does not take into account the problem of I/O congestion. 

Other efforts addressed I/O congestion as a global sche-\\duling problem using global I/O aware schedulers \cite{cars}, \cite{bbenabled}, \cite{zhou}. The goal of this approach is to optimize whole system utilization by minimizing I/O interference between different applications running on the system. However, this direction does not offer programming support to express opportunities of I/O performance improvements that are inherent in I/O intensive applications. Moreover, it focuses on optimizing system-wide performance metrics when handling the I/O of different running applications instead of optimizing the total performance of applications. 
   

In this work, we address the lack of support for mitigating the I/O performance bottleneck in task-based programming models. We argue that task-based programming models offer a suitable abstraction that can be leveraged to exploit parallelism opportunities in I/O intensive applications to improve total performance. On the one hand, I/O workloads can be wrapped by tasks whose execution overlap with the execution of compute tasks. Hence, application parallelism is increased. Furthermore, fine-grained I/O libraries can be still used for I/O optimization inside tasks \cite{nativempi}. On the other hand, task-based programming models have application-level information such as the number of I/O tasks and the I/O bandwidth requirement of each task, that can be used to manage I/O congestion.


More specifically, in this paper we propose enabling \textit{I/O Awareness} in task-based programming models. The main objective of an I/O aware task-based programming model is to improve the performance of I/O intensive applications by exploiting their inherent performance improvement opportunities. To this end, I/O aware task-based systems should support the following capabilities:
\begin{itemize}
    \item First, increasing task parallelism by defining \textit{I/O Tasks} to handle I/O workloads execution. I/O tasks execution can be overlapped with compute tasks execution.
    
    \item Second, managing I/O congestion by controlling I/O tasks scheduling through constraining tasks execution. 
\end{itemize}


Our proposal is realized by implementing the aforementioned I/O awareness capabilities in the PyCOMPSs task-based programming model \cite{pycompss}. I/O aware PyCOMPSs allows users to set constraints to I/O tasks to control their scheduling. Since the value of this constraint is fixed for the whole application execution, we call this a \textit{Static} constraint.

However, identifying a suitable constraint at application development time may be complex due to the lack of information about the amount of I/O that the application will produce and I/O performance on a given infrastructure. Therefore, we propose an automatic and abstract constraint mechanism that is exposed by the execution manager to settle on and tune the constraints of I/O tasks during application's execution. Hence, offering greater flexibility and portability. This mechanism carries out a performance exploration process to identify the optimal manner to execute I/O tasks on a given system. We call this type of constraints: \textit{Auto-Tunable} constraint. Using auto-tunable constraints, the burden of identifying the optimal constraint is removed by making the runtime system automatically infer and tune I/O tasks' bandwidth constraints with the goal of achieving total time performance benefit.

The main contributions of this paper can be listed as follows: 
\begin{enumerate}
    \item Programming model extensions to define I/O tasks and the runtime support to enable overlapping the execution of I/O tasks and compute task. This will increase the task parallelism and the application performance.

    \item Introducing I/O bandwidth constraints in the task definition and its runtime scheduling support to limit the number of concurrent I/O tasks and reduce I/O congestion.
    
    \item A methodology to automatically infer I/O bandwidth constraints at runtime in order to minimize the execution time of I/O tasks.
    
    \item A prototype implementation of these I/O awareness capabilities within the PyCOMPSs programming mo-\\del.  
\end{enumerate}

The prototype has been validated by applying these capabilities in a set of applications with different I/O workloads on the MareNostrum 4 supercomputer. We have compared their execution with a version that is not using the I/O awareness capabilities. The results show an improvement in the total application performance that can reach up to 43\%.

The rest of the paper is structured as follows. Section \ref{sec:related_work} discusses related work. Section \ref{sec:io_awareness} presents the main concepts and capabilities of an I/O aware task-based system. Section \ref{sec:implementation} gives an overview of the PyCOMPSs programming model and presents the design and implementation of the I/O awareness capabilities in PyCOMPSs: I/O tasks and storage bandwidth constraints. Section \ref{sec:evaluation} analyzes the performance results of I/O aware PyCOMPSs on the MareNostrum 4 supercomputer using different I/O workloads. Finally, Section \ref{sec:conclusion} concludes the paper and describes future work.

~\

\section{Related Work}
\label{sec:related_work}

Task-based programming models have gained popularity in recent years for orchestrating and executing Python applications \cite{efforts}. For instance, Parsl \cite{parsl} uses function decorators to compose workflows. Parsl provides a different set of extensible executors to address different parallelization requirements of applications and enable execution on different platforms.

Luigi \cite{Luigi} enables the explicit specification of dependency graphs. Using Luigi, users have to use the provided object oriented API to explicitly define dependencies in the code rather than annotating functions. At runtime, Luigi builds the execution graph by inspecting defined dependencies. 


The aforementioned frameworks and libraries target applications performance improvement through parallelization of computation and execution on variety of platforms. However, they do not offer any support that is specific for optimizing I/O performance nor addressing I/O performance bottlenecks. Unlike I/O aware PyCOMPSs, they do not have the notion of I/O tasks and there is no support for I/O-compute tasks overlap. Moreover, there is no programming model support for addressing I/O congestion.

The Dask {\cite{dask}} Python library implements parallel versions of Python libraries such as Numpy {\cite{numpy}} and Pandas {\cite{pandas}}. Dask enables specifying constraints on tasks execution which is similar to one of our contributions. Additionally, it is possible to explicitly specify zero CPU requirements for tasks which can allow for overlapping I/O and computation. However, the Dask runtime does not provide an automatic mechanism for setting and tuning constraints such as the one proposed in this paper.

Addressing the I/O performance bottleneck, numerous studies have been carried out at different levels. A traditional approach to improve application's I/O performance is to use I/O libraries, such as MPI-IO \cite{mpiio}, HDF5 \cite{hdf5} and NetCDF \cite{netcdf}. These libraries provide a programming API to manipulate data access. Therefore, applications do not need to assume the POSIX interface. MPI-IO provides a low-level interface to enable parallel I/O. This interface can be used to define how to access a file system to perform parallel I/O operations. On the other hand, HDF5 and NetCDF provide file formats that optimize the storage of large amounts of data by stipulating their formats and performing low-level optimizations. 



Using I/O libraries allows for fine-grained I/O optimization related to I/O access and storage. However, this approach does not address the problem of I/O congestion. Another limitation of this approach is portability: once these libraries are used in an application for a specific platform, it is not a straightforward task to use it on other platforms. 

More recently, I/O congestion has received a lot of research attention. Efforts in this direction addressed I/O congestion as a classical scheduling problem with the goal of optimizing for system-wide performance metrics. Gainaru et al. \cite{iocongestion} proposed a global I/O scheduler that has global view of the system and of the past behaviour of all applications running on it. These information can be used to optimize scheduler heuristics such as maximum efficiency or fairness. 

Liang et al. \cite{cars} proposed a contention-aware resource scheduling strategy to improve the performance of burst \\buffers by minimizing I/O congestion caused by I/O of different applications. This strategy analyzes I/O load on the burst buffers nodes and assigns incoming I/O to burst buffer nodes with least I/O load.  Herbein et al. \cite{bbenabled} incorporated I/O workload scheduling into existing policies such as First Come First Served (FCFS) and EASY backfilling. The idea is to add I/O as additional constraint when determining if a job can be scheduled. Jobs are only scheduled for execution if and only if there are available resources to satisfy their I/O requirements. 

Zhou et al. \cite{zhou} presented an I/O batch scheduler with two policies: conservative and adaptive. The conservative policy avoids I/O congestion as much as possible targeting system-performance metrics. Whereas the adaptive policy allows I/O congestion to happen to increase jobs performance. 


Unlike global I/O schedulers, our proposal does not target optimizing any system-wide performance metric, nor does it have any information about any other applications running on the system. I/O aware PyCOMPSs addresses I/O congestion from the view point of the application to increase its total performance. We consider that the previously mentioned efforts targeting I/O performance improvement (i.e. I/O libraries and I/O schedulers) can jointly work along with our proposal to achieve optimal application performance while improving the system-wide performance targets. 

Tillenius et~al. \cite{larsson} proposed a predictive model for task performance degradation and a resource aware scheduling policy by enabling users to set constraints for tasks execution using task annotation, similar in spirit to one of our contributions. 


\section{I/O Awareness in Task-based Programming Models}
\label{sec:io_awareness}


~\

\begin{figure}[htbp]
    \centering
    \includegraphics[width=\linewidth]{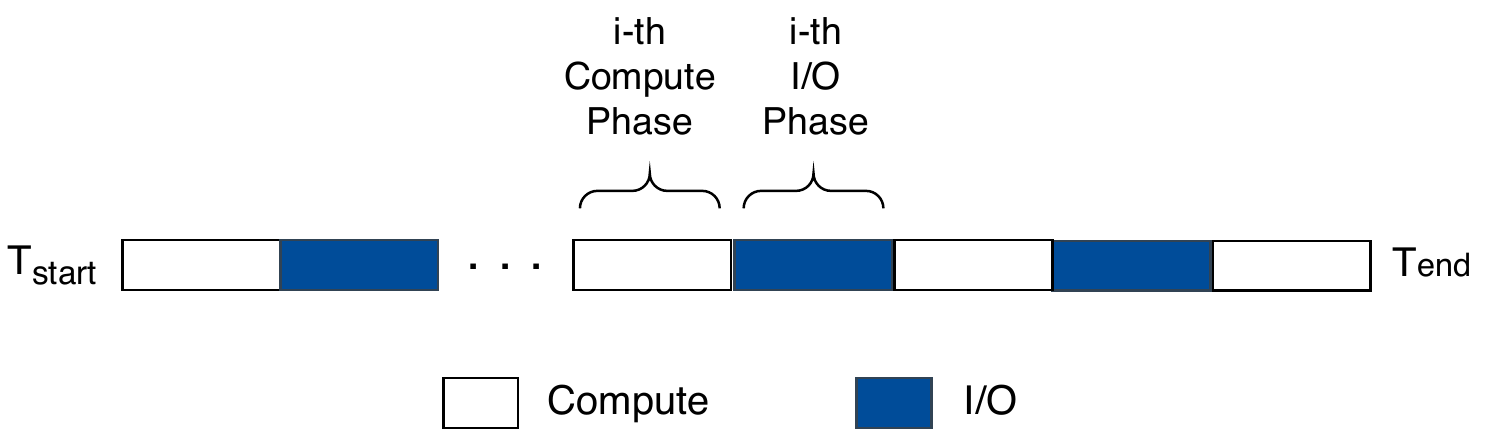}
    \caption{Life cycle of an I/O intensive application}%
    \label{fig:lifecycle}%
\end{figure}

A typical I/O intensive application consists of two phases: a computation phase and an I/O phase. Figure \ref{fig:lifecycle} shows a high-level abstraction of the life cycle of such applications. Each compute phase is followed by an I/O phase (e.g. checkpointing the results of the previous compute phase). Once each I/O phase is over, it is followed by another compute phase except at the end of application's execution where there is no more computation. The time of each I/O phase varies depending on the size of I/O  workload in each phase. It should be noted that this model assumes that the data consumed by the i-th I/O phase cannot be invalidated by the i+1 compute phase. This assumption is valid if the i+1 compute phase is independent from previous compute phases. Otherwise, if there is a data dependency between two compute phases (e.g. the i-th and i+1 compute phase), then the i+1 compute phase should receive an independent copy of the data so as to not invalidate the data consumed by i-th I/O phase.

Taking a closer look at Figure \ref{fig:lifecycle}, one can observe optimization opportunities that are possible due to the compute-I/O workloads pattern. For instance, the start of each compute phase (i+1) is delayed until the previous I/O phase (i) has finished. However, since there is no dependency between each I/O phase and the following compute phase, their execution can be overlapped. 

\begin{figure}[htbp]
    \centering
    \includegraphics[width=\linewidth]{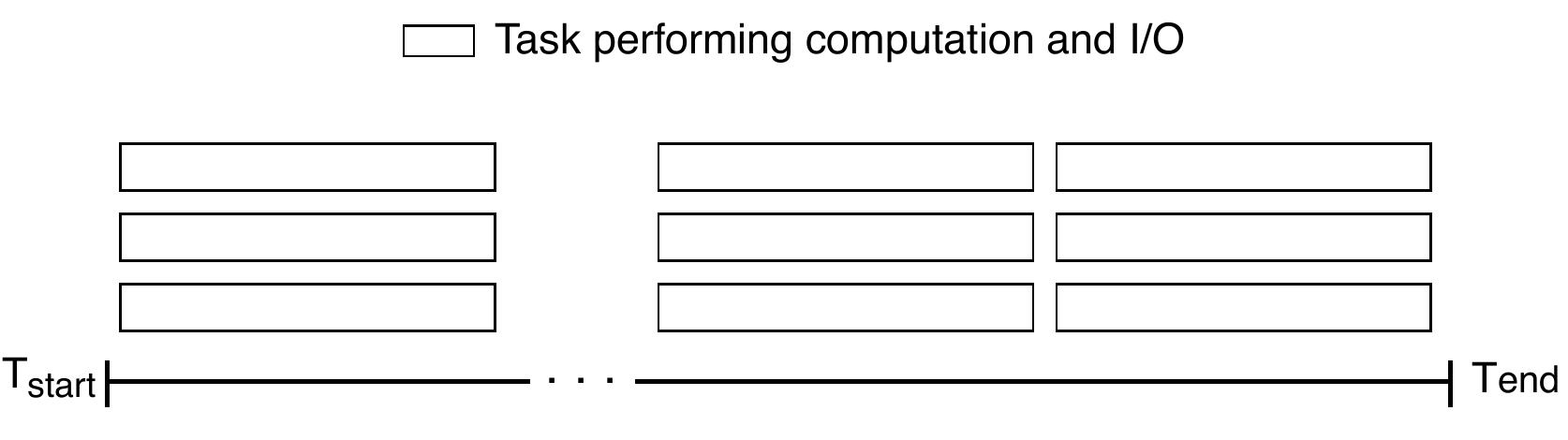}
    \caption{I/O intensive application executed with a traditional task-based programming model}%
    \label{fig:traditional}%
\end{figure}

A traditional task-based system will not be able to exploit the parallelism opportunities of applications with such pattern. Figure \ref{fig:traditional} shows a high-level abstraction of how a traditional system can be used to execute an I/O intensive application. Computation and I/O are executed in one task without the ability to differentiate between each workload. Using such approach wastes a lot of performance improvement opportunities for both computation and I/O. This is due to two reasons: (i) application's parallelism is decreased because computing resources cannot execute any compute workload while they are waiting for I/O completion. (ii) the scheduling of workloads is limited; tasks scheduling can be optimized for either compute performance or I/O performance but not for both. For instance, launching more computing workloads in parallel usually results in more performance. However, in the case of I/O workloads it could result in increasing I/O congestion.

\begin{figure}[htbp]
    \centering
    \includegraphics[width=0.8\linewidth]{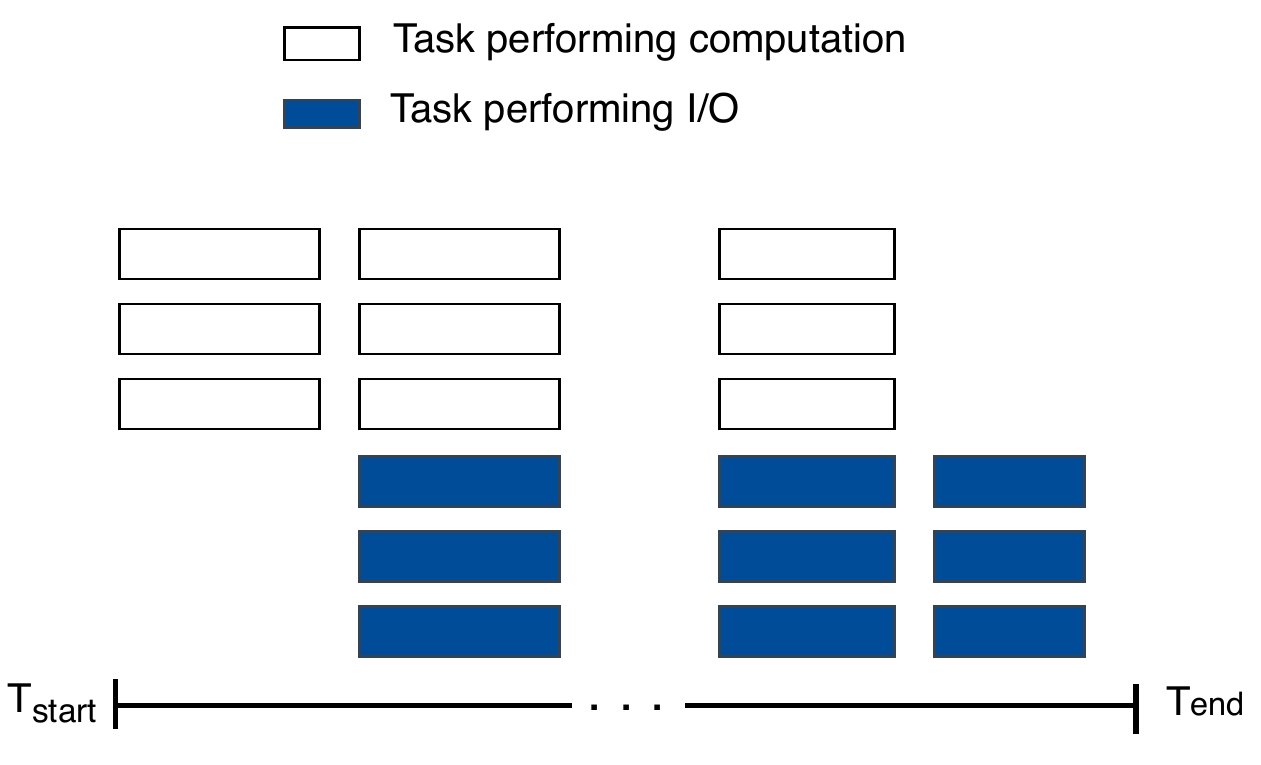}
    \caption{I/O intensive application executed with an I/O aware task-based programming model}%
    \label{fig:aware}%
\end{figure}

To take advantage of the performance improvement opportunities present in Figure \ref{fig:lifecycle}, we propose enabling \\ \textit{I/O awareness}. I/O Awareness enables task-based models to separate compute and I/O workloads. Therefore, allowing for the optimization of each workload depending on its properties. Figure \ref{fig:aware} shows the life cycle of an application executed with an I/O aware task-based model. This application has two types of tasks: tasks that execute compute workloads and tasks that execute I/O workloads. In an I/O aware execution, compute workloads execution can be overlapped with the execution of dependency free I/O workloads (i.e. the I/O workloads executed by tasks which their data dependencies are satisfied and can be released for execution). Thus, the level of parallelism is increased due to the overlapping execution of tasks. In addition to that, I/O workloads can be scheduled independently from compute workloads to improve I/O performance by minimizing I/O congestion.

Sections \ref{subsec:concepts_io_tasks} to \ref{subsec:concepts_auto_constraints} present the I/O awareness capabilities that we propose to be supported by task-based models: Section \ref{subsec:concepts_io_tasks} introduces the concept of \textit{I/O Tasks}, which allow to take advantage of the proprieties of I/O workload to increase task parallelism by overlapping tasks execution. Next, Section \ref{subsec:concepts_io_congestion} focuses on addressing I/O-specific performance problems such as I/O congestion by constraining tasks scheduling. Finally, Section \ref{subsec:concepts_auto_constraints} proposes the automatic inference of task constraints during application's execution.

~\

\subsection{I/O Tasks}
\label{subsec:concepts_io_tasks}

~\


We define \textit{I/O tasks} as special tasks for exclusively executing I/O workloads in applications. Indeed, I/O tasks can contain a single I/O request or several consecutive requests (e.g., a loop of write accesses). 

Unlike current task-based systems that schedule all tasks based on computing constraints, I/O aware systems should be able to schedule I/O tasks according to their I/O bandwidth requirements instead of computing requirements. \\Hence, I/O tasks scheduling will not be bounded by computing infrastructure nor by the compute workload in the application because their scheduling will not depend on the availability of computing units. This approach allows the scheduling of as many concurrent I/O tasks to reach the peak performance of the storage infrastructure. 

The execution of I/O tasks can be overlapped with the execution of compute tasks. Dependency-free I/O tasks can be executed along with compute tasks on the same CPU with negligible impact on the performance of compute tasks. This takes advantage that CPUs remain idle during I/O execution, waiting for the I/O request to be completed and the data to be transferred to the storage device. This capability increases task parallelism in applications because computing resources will not be occupied solely for executing I/O workloads.

Moreover, Using I/O tasks to identify I/O workloads enables the scheduling of I/O tasks to specialized storage subsystems in distributed heterogeneous infrastructures. These storage subsystems allow higher I/O performance because they are equipped with different storage hardware such as Solid State Drives (SSDs) or Non-Volatile Memories \\(NVMe) \cite{nvram} that offer higher bandwidth capabilities than traditional Hard Disk Drives (HDDs).


\subsection{Storage Bandwidth Constraints}
\label{subsec:concepts_io_congestion}

~\

The second capability that we propose to enable I/O \\awareness in task-based systems is the ability to address the problem of I/O congestion. I/O congestion mainly occurs because the aggregate amount of data to be written by the concurrently running I/O tasks surpasses the maximum I/O bandwidth of the storage devices. Consequently, assuming that the I/O bandwidth is fairly allocated between concurrent I/O tasks, the more I/O tasks running concurrently, the less I/O bandwidth would be allocated to serve the requirements of each task. Therefore, the execution time of I/O tasks increases leading to not only degraded I/O performance but also total performance degradation. Therefore, I/O awareness which only supports I/O tasks may not be enough to achieve total performance improvement.

I/O congestion can be tackled by constraining the scheduling of I/O tasks. Using constraints to control I/O tasks scheduling guarantees that only a maximum number of tasks can run concurrently at any given time of application's execution. Hence, I/O bandwidth can be managed and I/O congestion can be minimized or completely avoided. 

To this end, we propose enabling the use of \textit{Storage Bandwidth} constraints to specify an estimate for the I/O bandwidth requirements of a task. Using the storage bandwidth constraint implies that if scheduling a task will over-allocate the storage bandwidth of the available storage resources, then this task should not be scheduled even though compute resources are idle. Tasks with storage bandwidth constraint will only be scheduled if there is available I/O bandwidth to satisfy their constraints. Otherwise, they will wait for more storage bandwidth to become available.

One approach to specify the storage bandwidth constra-\\ints is to extend the programming model to allow users to set it at application development time. Hence, users can plan the execution of applications by controlling the level of I/O tasks parallelism that would benefit their applications on a given infrastructure.


\subsection{Automatic Inference of Storage Bandwidth Constraints}
\label{subsec:concepts_auto_constraints}

~\

Identifying a suitable storage bandwidth constraint that minimizes I/O congestion and improves I/O and total application performance may be difficult at application development time. Indeed, a high storage bandwidth constraint leads to a lower number of concurrent tasks and more bandwidth allocated to each task. Hence, I/O congestion will be minimized and I/O performance will increase but task parallelism will decrease.  Similarly, a low storage bandwidth constraint will allow more tasks to be executed concurrently but less I/O bandwidth will be allocated for each task. Hence, task parallelism will increase but increased I/O congestion will negatively affect application's performance.

To overcome the aforementioned difficulty, we propose that I/O aware systems support the automatic inference of storage bandwidth constraints. The main objective of this mechanism is to allow the runtime system to automatically estimate a constraint that is not very high so it allows more I/O tasks to run concurrently in order to maximize task parallelism. At the same time, this constraint should not be very low so it minimizes I/O congestion as much as possible by avoiding the launch of a lot of I/O tasks concurrently.

More specifically, the automatic inference of a constraint is the process of finding a constraint that maximizes task parallelism and minimizes I/O congestion. 

In order to identify such a constraint, we propose a two-steps mechanism: First, the runtime system runs a \textit{learning phase}, in which it collects information about the I/O tasks performance with different levels of I/O tasks parallelism (i.e. different constraint values). Second, the information collected during the learning phase is applied to a heuristic function with the objective of minimizing the execution time of the I/O tasks to be scheduled. 

Section \ref{subsubsec:concepts_learning} describes in more detail the learning phase, whereas Section \ref{subsubsec:concepts_min} presents the objective function for setting an optimal constraint given a number of I/O tasks.


\subsubsection{Learning Phase}
\label{subsubsec:concepts_learning}

~\


During the learning phase, the system keeps track of the average I/O task time when running different number of concurrent I/O tasks. Indeed, the number of concurrent I/O tasks at any moment of the application execution is controlled by the value of the constraint that is used. Therefore, the system tries different constraint values to launch different number of concurrent I/O tasks. 

The learning phase consists of several \textit{Learning Epochs}. In each learning epoch, the system uses a different constraint value to launch different number of concurrent I/O tasks. The purpose of each learning epoch is to identify the average I/O task time when using a certain constraint value. Therefore, a learning epoch can be defined as the set of I/O tasks that are allowed to run concurrently when using a certain constraint.  

It should be noted that the lifetime of an epoch is not defined by any time limits nor by any assumptions based on the task-graph properties. For example, if the maximum number of tasks allowed to run concurrently when using a certain constraint is 5, then the lifetime of the learning epoch of this constraint is the execution time of the 5 concurrent tasks. Once the average I/O task time of the maximum number of tasks allowed to run concurrently at any given time is obtained, a learning epoch is ended and the next epoch (where different constraint is used) is started.

Different approaches can be used to determine the details of the learning phase (i.e. the number of learning epochs and how to progress the learning phase). We propose two types of auto-tunable constraints: \textit{bounded} and \textit{unbounded}.

In the case of the bounded auto-tunable constraint, three values can be used to control the learning phase: \textit{minimum} and \textit{maximum} constraint values that set the boundaries of the constraint and a \textit{delta} value which represents the step size that allows the progression from the \textit{minimum} value to \textit{maximum} value. The first learning epoch in the learning phase starts with the \textit{minimum} constraint value, then the learning phase progresses until it reaches the \textit{maximum} constraint value by multiplying the current constraint by the value of \textit{delta}. 

However, in the case of the unbounded automatic constraint, no values are used to bound or guide the learning phase, instead such values are estimated by the runtime system. An unbounded auto-tunable constraint starts with the lowest constraint value that would allow the maximum number of I/O tasks to run concurrently. After each learning epoch, the constraint value is doubled and used as the constraint of the next learning epoch. Doubling the constraint would progress the constraint value without risking skipping possible optimal constraint values. 

After each learning epoch, the following condition is evaluated to decide whether to continue or end the learning phase:
\[\scalebox{1.2}{$t_{Epoch(i)}$} \leqslant \scalebox{1.2}{$t_{Epoch(i-1)}$}/2  \]
where:~\\
\indent \scalebox{1.2}{$t_{Epoch(i)}$}:  average execution time of I/O tasks in learning epoch $i$

~\\

This condition assumes that since the constraint is doubled each new learning epoch (i.e. the number of concurrent tasks is halved), then the average task time in a learning epoch should decrease by at least half compared to the average task time recorded in the previous learning epoch.

Comparing the bounded and unbounded auto constraints, the bounded auto constraint has the ability to achieve more fine-grained results. This is because it has a longer learning phase where it tries as high constraint as the \textit{maximum} constraint value set by the user. Whereas the unbounded auto constraint follows a stricter learning approach. It follows the assumption that not getting the expected I/O task time improvement in a learning epoch will lead to a divergence path where no more improvement should be expected. Indeed, these different behaviours lead to different application performances.

More details on how to acquire and calculate the hyperparameters of the learning phase for the bounded constraint (i.e. \textit{minimum}, \textit{maximum} and \textit{delta}) and the starting constraint value for the unbounded constraint are described in the implementation section (Section \ref{subsubsec:impl_learning}).

~\\

\subsubsection{Objective Function}
\label{subsubsec:concepts_min}

~\

After the learning phase ends, the information that has been collected about the average task time when launching a certain number of concurrent tasks (i.e. using a certain constraint) are applied to minimize the following objective function: 
\begin{equation} \label{eq:fn}
 \scalebox{1.1}{$\forall c\in C\!\!:$} \quad \scalebox{1.1}{$\min$} \;\; \scalebox{1.1}{$\mathrm{T}(numTasks, \mathrm{c})$}
\end{equation}

\noindent where:~\\
\indent $C$: is the set of constraints used during the learning phase.

\indent $T(numTasks, c)$: is the time estimation for executing the given number of I/O auto-constrained tasks using the constraint $c$. This function can defined as:

 \[\scalebox{1.1}{$\mathrm{T}({numTasks, c}) =$}\] 
 \[\quad\quad\quad \scalebox{1.1}{$(numTasks/maxNumTasks_c) * \mathrm{t_c}$} \]

\noindent where:~\\
\indent $maxNumTasks_c$: is the maximum number of concurrent I/O tasks allowed to run using the given constraint $c$.

\indent $t_c$: is the average I/O task time when using the given constraint $c$. 

~\

The objective of this function is to choose a constraint that minimizes the execution time of the auto-constrained tasks waiting to be scheduled. In this function, the number of execution groups in which the tasks will be executed is calculated by dividing the given number of tasks by the maximum number of concurrent tasks using a given constraint $c$. Then this number is multiplied by the average task time obtained during the learning phase. 

After evaluating the objective function with all the constraints used during the learning phase, the constraint that results in the least execution time is assigned to the task.


\section{Implementation}
\label{sec:implementation}

~\

This section describes the PyCOMPSs framework and the implementation details of the proposed I/O awareness capabilities. Section \ref{sec:background} starts by giving a general overview of the PyCOMPSs programming model and runtime system. Next, Section \ref{subsec:io_aware_pycompss} describes some implementation details about the I/O awareness capabilities in PyCOMPSs. 

\subsection{PyCOMPSs Overview}
\label{sec:background}

~\

PyCOMPSs is a Task-based programming model that enables the parallel execution of Python applications in a task-based manner on distributed infrastructures. PyCOMPSs relies on the COMPSs framework \cite{compss} to exploit application parallelism at task level and manage the execution of tasks on various distributed infrastructures such as grids, large clusters and clouds. Section \ref{subsubsec:prog_model} describes the programming semantics of PyCOMPSs. Whereas Section \ref{subsubsec:runtime} gives an overview about the COMPSs runtime.

\subsubsection{PyCOMPSs Programming Semantics}
\label{subsubsec:prog_model}

~\

The objective of PyCOMPSs is to allow distributed execution of applications without compromising programmability and ease of development. Therefore, the programming model of PyCOMPSs allows converting sequential applications to task-parallel applications with minimal code modification. Using PyCOMPSs, a functions can be defined as a task by annotating it with the \textsourcecode{@task} Python decorator.

The code snippet in Listing ~\ref{listing:pycompss_sample_code} shows an example of a PyCOMPSs application. This example has two functions that are declared as tasks: (i) \verb|scale| (line 2) and (ii) \verb|accumulate| (line 7). In the \verb|@task| decorator, users have to specify the return types of the task outputs -if any- and the directionality of the task parameters. For each of the task parameters, the directionality of a certain parameter describes whether this parameter will be read (\verb|IN|), updated (\verb|INOUT|) or written (\verb|OUT|). The directionality parameters are used later by the runtime to identify dependencies between tasks. 

In Listing ~\ref{listing:pycompss_sample_code}, the \verb|scale| task is invoked for each \verb|INT| in the \verb|inputs| list. Then, the \verb|accumulate| task is called to sum the scaled values where it updates the parameter \verb|value1|. Since PyCOMPSs tasks return future objects in order to allow asynchronous execution, a synchronization point is defined at line 16 (\verb|compss_wait_on|) to request the final output value.   

In addition to the \verb|@task| decorator, PyCOMPSs provides other decorators such as the \verb|@constraint| decorator to enforce hardware or software requirements on task execution. For instance, the \verb|computingUnits| constraint specifies a required number of CPUs for a task. A constrained task will be launched for execution only if its constraints are satisfied. In line 5 of Listing ~\ref{listing:pycompss_sample_code}, a constraint of 2 computing units is set as a requirement for the execution of the \verb|accumulate| task.

\begin{listing}[!htb]
  \centering
  \begin{minted}[xleftmargin=0.7cm, linenos]{python}
@task(returns=int)
def scale(c):
    return c * 10

@constraint(computingUnits=2)
@task(value1=INOUT)
def accumulate(value1, value2):
    value1 += value2

if __NAME__ == "__main__":
    inputs = [INT1, INT2, .., INTn]
    result = 0
    for num in d:
        scaled = scale(num)
        accumulate(result, scaled)
    result = compss_wait_on(result)
    \end{minted}
\caption{PyCOMPSs Sample Code}
\label{listing:pycompss_sample_code}
\end{listing}

\subsubsection{The COMPSs Runtime}
\label{subsubsec:runtime}

~\

The COMPSs runtime is responsible for detecting data dependencies between tasks and managing their execution. To allow execution on distributed infrastructure, the deployment of the COMPSs runtime follows a master-worker \\paradigm: one node acts as the \textit{master} node and the rest of the nodes act as \textit{worker} nodes. 

\begin{figure}[htbp]
    \centering
    \includegraphics[width=0.9\linewidth]{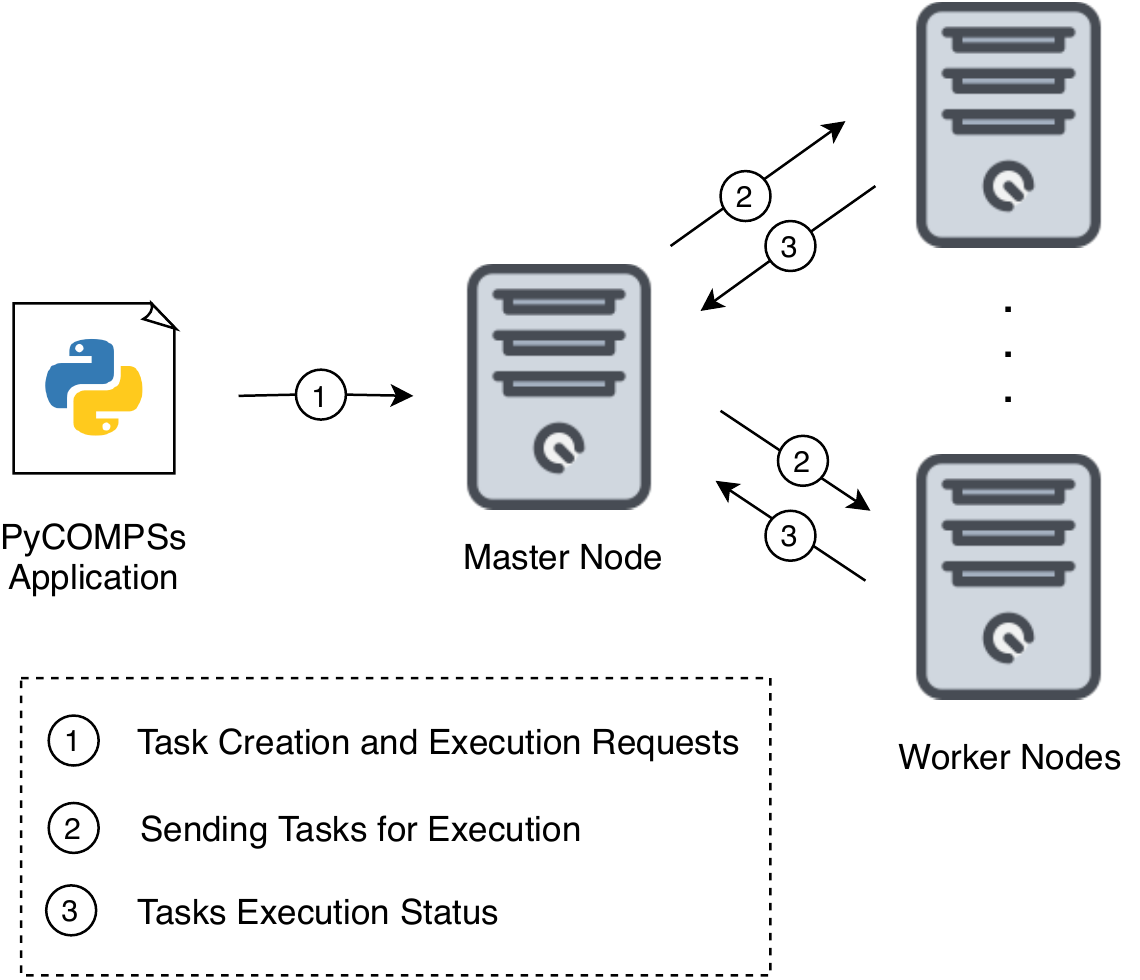}
    \caption{Overview of the COMPSs Runtime}%
    \label{fig:master}
\end{figure}

Figure \ref{fig:master} depicts a high-level overview of the COMPSs runtime during application execution. After launching the application, the COMPSs master component on the master node starts to receive tasks creation and execution requests. After analyzing data dependencies between the tasks, the COMPSs scheduler assigns dependency-free tasks to one of the workers to be executed. When tasks execution end, the worker component of the worker node notifies the master with the execution status.

\begin{figure}[htbp]
    \centering
    \includegraphics[width=\linewidth]{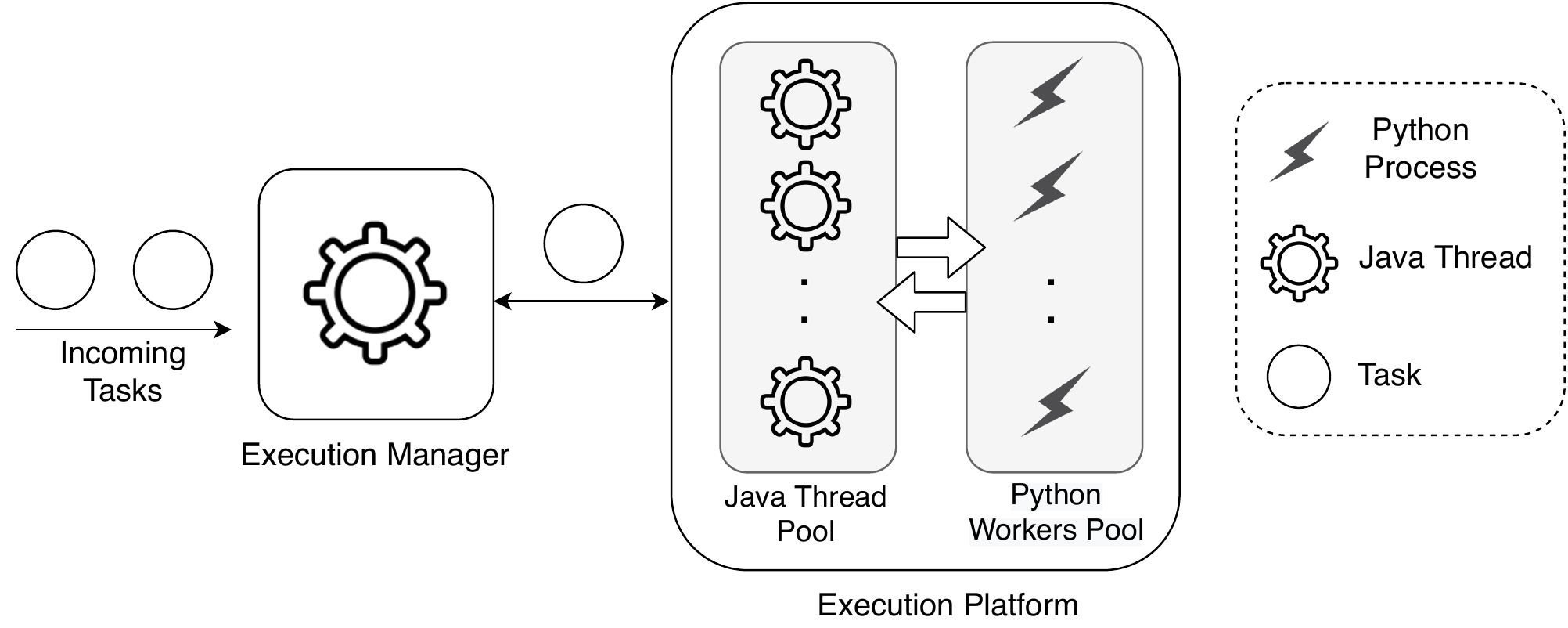}
    \caption{COMPSs Worker Component}%
    \label{fig:worker}
\end{figure}

Figure \ref{fig:worker} depicts an overview of the architecture of the worker component of the COMPSs runtime. Every worker component consists of a Java process called the \textit{Execution Manager} which is responsible for setting up and managing an execution platform at the launch time of the application. The execution platform consists of a Java thread pool with as many threads as the number of CPUs on the worker. These Java threads are called \textit{Executors}. Each Java executor handles the execution of one task. Furthermore, each Java executor launches a Python process, called \textit{Python Worker}, that will ultimately carry out the execution of the task. Java executors and Python workers communicate with each other using operating system pipes. In order to facilitate the communication between the Java and Python components of the system and data transfer between nodes, tasks outputs are stored in a serialized format.

In order for the COMPSs scheduler to make scheduling decisions that honours task constraints, the COMPSs runtime allows users to describe the available execution resources by providing a \textit{Resource Description} file in an XML formal. 

During tasks execution, the COMPSs runtime keeps track of how much resources are consumed by currently running tasks and how much resources are free. When a task arrives for scheduling, the COMPSs scheduler compares the constraints/requirements of this task to the available resources at the moment of task arrival to decide if there are enough resources to launch the task for execution.

\subsection{I/O Awareness in PyCOMPSs}
\label{subsec:io_aware_pycompss}

~\

In this section we present implementation details of the I/O awarness capabilities in PyCOMPSs. Section \ref{subsec:io_tasks} \\presents the special handling of the I/O workload through the use of \textit{I/O Tasks} and how can their execution be overlapped with the execution of compute tasks. Section \ref{subsec:io_congestion} describes how task constraints are used to control I/O tasks scheduling to minimize I/O congestion. Finally, Section \ref{subsubsec:auto_constraints} presents the details of the automatic storage bandwidth constraint inference in PyCOMPSs.

\subsubsection{I/O Tasks} 
\label{subsec:io_tasks} 

~\
 
Following task declaration conventions of PyCOMPSs, a task is declared as I/O task by the means of the \verb|@IO| decorator. Listing \ref{listing:iotasks} shows the I/O tasks annotation in PyCOMPSs. Besides using the \verb|@task| decorator to define a PyCOMPSs task, the \verb|@IO| decorator is used to declare that this task should be handled as an I/O task. 


\begin{listing}[!htb]
  \centering
  \begin{minted}[xleftmargin=0.7cm, linenos]{python}

@IO()
@task()
def io_task(data):
    # perform I/O operations on data 
    
    \end{minted}
\caption{I/O Task Annotation}
\label{listing:iotasks}
\end{listing}

\begin{figure}[htbp]
    \centering
    \includegraphics[width=\linewidth]{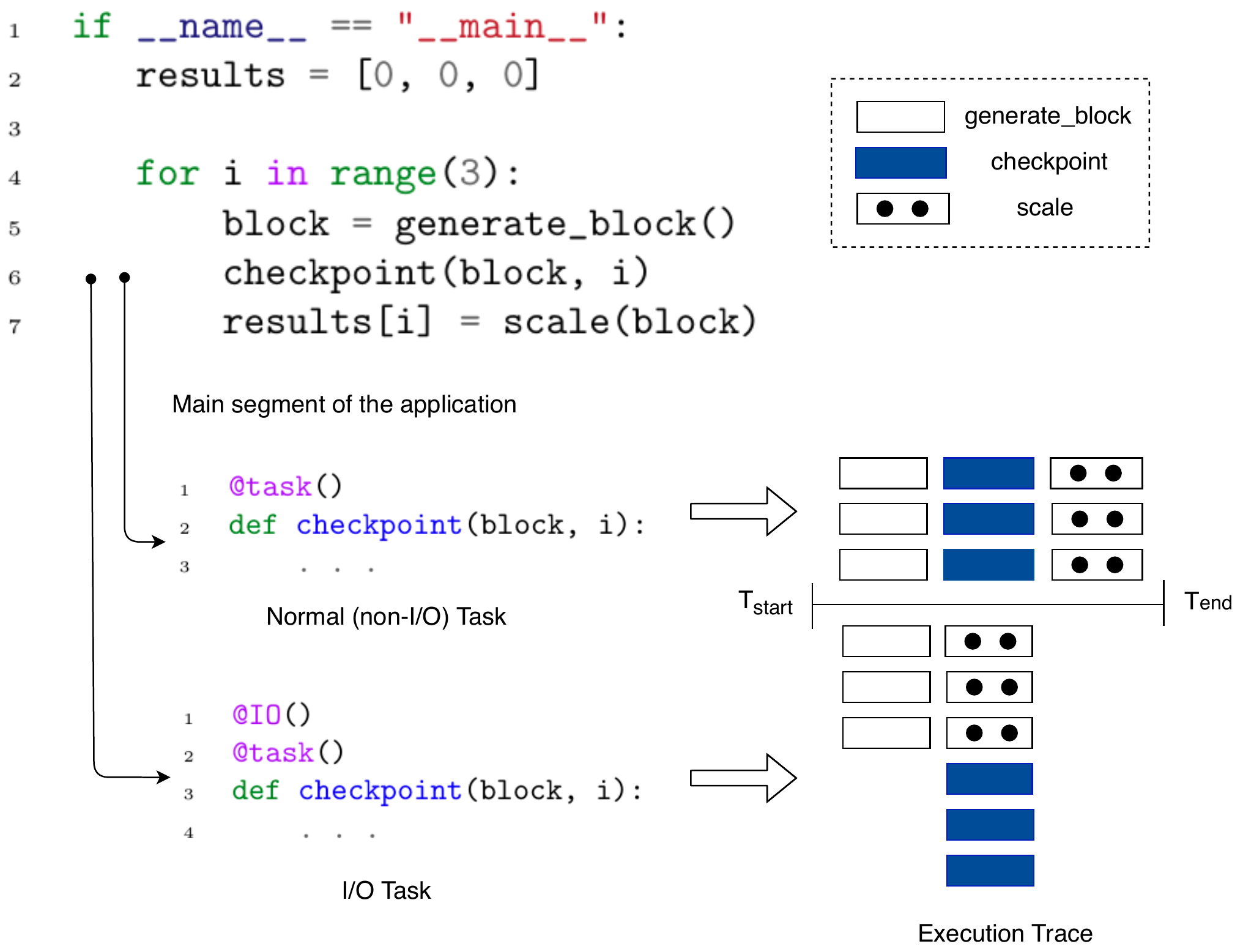}
    \caption{I/O tasks overlap with compute tasks}%
    \label{fig:overlap}%
\end{figure}

Figure \ref{fig:overlap} shows how using I/O tasks affect the execution. In the main code snippet, a loop launches three tasks: (i) \verb|generate_block| task which returns a block of a certain size. (ii) \verb|checkpoint| task which writes the block to the disk. (iii) \verb|scale| task which does some computation on the block, the output of this task is stored in the \verb|results| list. Both \verb|checkpoint| and \verb|scale| tasks are dependent on the \\ \verb|generate_block| task, however, they do not have dependencies between each other. Therefore, their execution can overlap. 

As shown in Figure \ref{fig:overlap}, the \verb|checkpoint| task can be handled in two different ways during the execution of the application depending on how it is defined in the code: On the one hand, it can be defined as a normal task by only using the \verb|@task| decorator. Consequently, the execution of the \verb|scale| tasks will be delayed until the \verb|checkpoint| tasks finish execution. On the other hand, the \verb|checkpoint| task can be defined as an I/O task by using the \verb|@IO| decorator. This way, the COMPSs runtime handles the \verb|checkpoint| tasks as I/O tasks, hence, the \verb|scale| compute tasks are launched and the execution of both tasks is overlapped.

In order to enable the overlapping execution between dependency free I/O tasks and compute tasks like in Figure \ref{fig:overlap}, modifications were made to the master and worker components of the COMPSs runtime. In the master component, by default, the COMPSs runtime assigns one CPU to every task. Consequently, the runtime scheduler will not launch any new tasks for execution unless there is enough CPUs to execute the task. However, unlike regular compute tasks, when the COMPSs runtime receives an I/O task creation request, it sets its computing requirements to zero. Consequently, incoming I/O tasks will be scheduled immediately even if all the CPUs in the infrastructure are consumed by compute tasks. 

In the case a shared working directory between tasks is specified, it will be used to store tasks outputs in a serialized format. Therefore, no node-to-node data transfer is required and, as a consequence, I/O tasks are scheduled to the first candidate node. However, if the working directory is not shared, then I/O tasks will be scheduled taking into consideration data locality.



Indeed, for the case when a shared working directory is used, alternative design approaches can be adopted to distribute the tasks to candidate workers (e.g. round-robin). However, the aforementioned behavior provides the most general approach and least intrusive implementation. In addition to that, our focus in this paper is to schedule I/O tasks based on their bandwidth requirements using task constraints instead of imposing a certain scheduling policy.

As for the worker component of the runtime, it needs to support CPU oversubscription to enable the execution of I/O tasks side by side with compute tasks on the same CPU. Therefore, we added an execution platform, called \textit{I/O Execution Platform}, dedicated to handling I/O tasks execution while the other execution platform, called \textit{Compute Execution Platform}, is dedicated to handling compute tasks execution. 

\begin{figure}[htbp]
    \centering
    \includegraphics[width=0.9\linewidth]{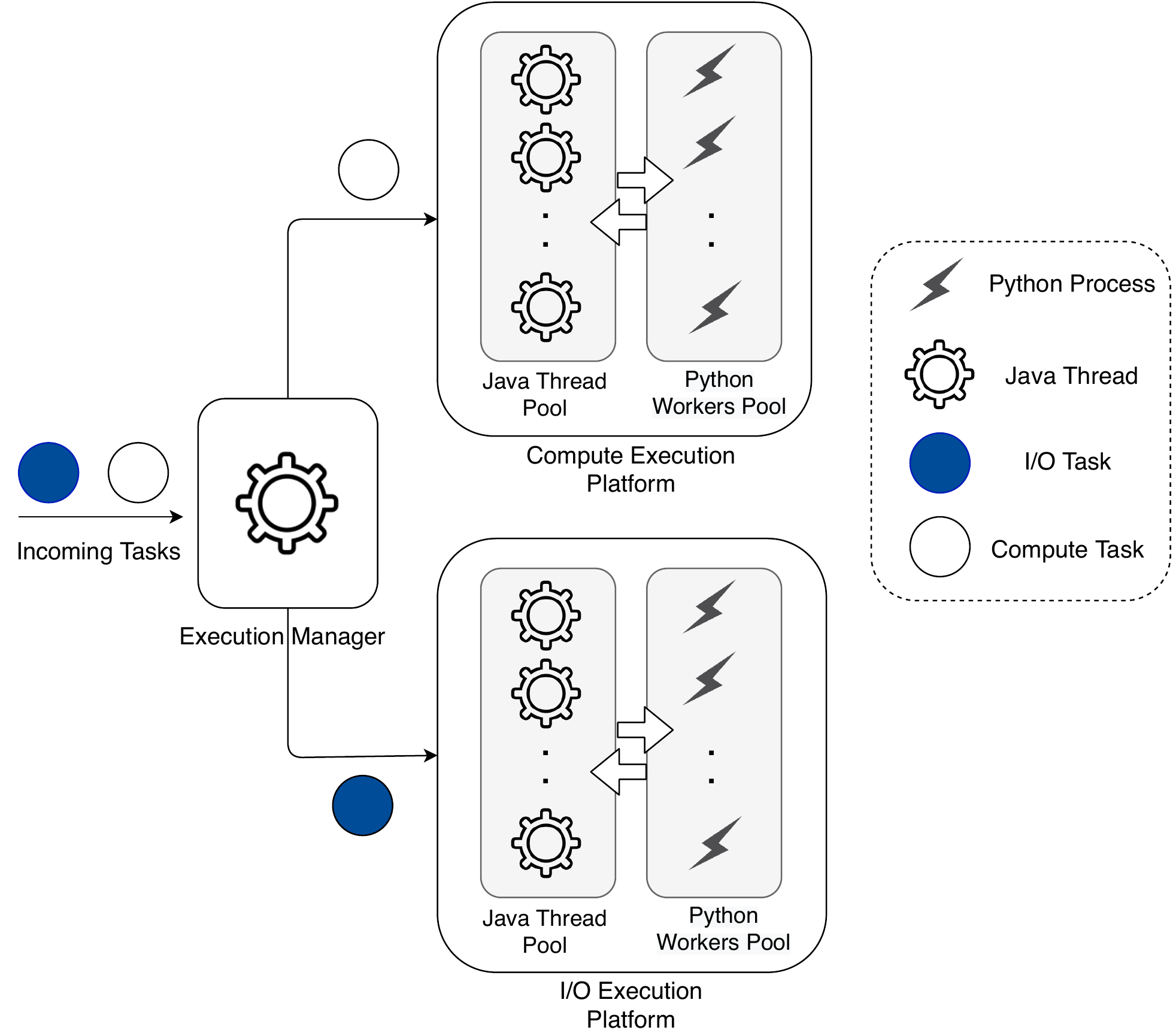}
    \caption{I/O Aware PyCOMPSs Worker; I/O Execution Platforms handles the execution of I/O tasks}%
    \label{fig:worker_ioaware}%
\end{figure}

Figure \ref{fig:worker_ioaware} illustrates a high-level overview of the architecture of the I/O aware worker component. Similar to the compute execution platform, the I/O execution platform handles the execution of I/O tasks by managing a number of executor threads. These executor threads are created at the launch time of the application and their number can be set in the PyCOMPSs launch command. 

\subsubsection{Static Storage Bandwidth Constraints}
\label{subsec:io_congestion}

~\

In order to constrain I/O scheduling to avoid I/O congestion, we extended the \verb|@constraint| decorator to support a storage bandwidth constraint. During application execution, the COMPSs runtime keeps track of the state of the available I/O bandwidth of the system by updating its value based on the storage bandwidth requirement of each task.   

Listing \ref{listing:static_constraint} shows a sample I/O task with the storage bandwidth constraint. Using the \verb|storageBW| argument of the \\ \verb|@constraint| decorator, users can set an estimated bandwidth for I/O tasks. During the scheduling of the \\ \verb|constrained_write| task, the PyCOMPSs scheduler will use its storage bandwidth constraint value to determine when this task should be scheduled. If the storage bandwidth constraint of the  \verb|constrained_write| task is not satisfiable, then its execution will be blocked until its requirement becomes available.    

\begin{listing}[!htb]
  \centering
  \begin{minted}[xleftmargin=0.7cm, linenos]{python}

@constraint(storageBW = 20)
@IO()
@task()
def constrained_write(data):
    ... 
    
    \end{minted}
\caption{Constrained I/O task using storage bandwidth constraint}
\label{listing:static_constraint}
\end{listing}

The storage bandwidth constraint presented in Listing \ref{listing:static_constraint} is a \textit{Static} constraint. This means that the value of the constraint is set before launching the application. Moreover, the value of this constraint does not change during the execution of the application.

In order to allow the COMPSs scheduler to reason about the available storage bandwidth in the system, we extended the resources description file to enable users to specify the maximum I/O bandwidth of the storage devices. This value will be used by the COMPSs runtime during applications execution to make better scheduling decisions. The currently supported unit to specify the required bandwidth for tasks in the @constraint decorator is MB/second.



\subsubsection{Auto-tunable Storage Bandwidth Constraints}
\label{subsubsec:auto_constraints}

~\

In this section, we present how auto-tunable constraints can be specified programmatically in PyCOMPSs, then we discuss more details about the learning phase and the objective function. For brevity, we will refer to \textit{Auto-tunable Storage Bandwidth Constraints} as \textit{Auto Constraints} in the rest of this paper.    

~\

A) \textit{Auto Constraints Syntax}
~\\

As discussed in Section \ref{subsec:concepts_auto_constraints}, we enabled two different types of auto constraints: \textit{Bounded} and \textit{Unbounded}. On the one hand, users can set bounded auto constraint as \textit{auto(min, max, delta)}; where \textit{min} represents the minimum starting constraint, \textit{max} represents the maximum possible constraint and \textit{delta} represents the value by which the runtime advances the constraint value from \textit{min} to \textit{max}. On the other hand, users can specify an unbounded auto constraint by setting the value of the storage bandwidth constraint to \textit{auto}. 

Listing \ref{listing:auto_bounded_constraints} shows an example of bounded auto constraint. To set a bounded auto constraint, users have to estimate the hyper-parameters that will guide the learning phase (i.e. min, max and delta). Whereas Listing \ref{listing:auto_unbounded_constraints} shows an unbounded auto constraint, in which the runtime will estimate the min, max and delta hyper-parameters.       

\begin{listing}[!htb]
  \centering
  \begin{minted}[xleftmargin=0.7cm, linenos]{python}

@constraint(storageBW = "auto(10,50,4)")
@IO()
@task()
def constrained_io_task(data):
    ... 
    
    \end{minted}
\caption{Bounded Automatic Constraint with Syntax \textit{auto(min, max, delta)}.}
\label{listing:auto_bounded_constraints}
\end{listing}

\begin{listing}[!htb]
  \centering
  \begin{minted}[xleftmargin=0.7cm, linenos]{python}

@constraint(storageBW = "auto")
@IO()
@task()
def constrained_io_task(data):
    ... 
    
    \end{minted}
\caption{Unbounded Automatic Constraint.}
\label{listing:auto_unbounded_constraints}
\end{listing}

B) \textit{The Learning Phase}
\label{subsubsec:impl_learning}
~\\

The COMPSs runtime automatically estimates the auto-constraints by carrying out the two-steps mechanism discussed in Section \ref{subsec:concepts_auto_constraints}. After the completion of the learning phase, the runtime will have an  \textit{Auto Constraint Registry} for each auto-constrained task. This auto constraint registry contains pairs of: constraint value and the average I/O task time when using this constraint \textit{(\{Constraint $\rightarrow$ Avg I/O Task Time\})}.

It should be noted that defining an auto constraint for a certain task will not affect how the runtime handles the other tasks in the application. In addition to that, it is possible to have different auto-constrained tasks in the same application. The COMPSs runtime will run a separate learning phase for each auto-constrained task and will set a constraint suitable for the workload of each task. We assume that an I/O task will always produce the same I/O workload during the application lifetime.

In the case of an unbounded automatic constraint, the runtime calculates the value of the starting constraint by dividing the maximum I/O bandwidth of the storage device by the number of I/O executors in each worker node. The number of I/O executors is a convenient choice for calculating the starting constraint because it represents the maximum number of I/O tasks that can run concurrently during any time of application execution.

In order to guarantee the integrity of the learning phase, the scheduler dedicates a worker node for each auto-constr-\\ained task in an active learning phase. These nodes are called \textit{Active Learning Nodes}. Once a node is marked as an active learning node for a specific auto-constrained task, the scheduler avoids scheduling any other I/O tasks or auto-constrain-\\ed I/O tasks to that node. Therefore, it is guaranteed that the learning phase of an auto-constrained task will not be interfered by the other I/O tasks. As soon as the learning phase of an auto-constrained task ends, the scheduler un-marks the associated active learning node and use it for scheduling as normal. It should be noted that the compute tasks are scheduled normally on all available nodes because they do not consume any storage bandwidth resources.

We study the performance of both types of automatic constraint and the effect of changing their hyper-parameters (the values of \textit{max, min, delta} in the bounded constraint and the number of I/O executors per worker node in the unbou-\\nded constraint) in the evaluation section (Section \ref{sec:evaluation}).

~\\ 
C) \textit{The Objective Function}
~\\

After finishing the learning phase for a certain auto-\\constrained task, the runtime applies the \textit{auto constraint registry} to objective function \ref{eq:fn} to choose a constraint that will result in the execution of the pending auto-constrained tasks in the least possible time. 

Some cases are considered when evaluating objective \\function \ref{eq:fn} for a given number of scheduling-ready auto-\\constrained tasks:
\begin{itemize}
    \item If the number of tasks is not divisible by the maximum number of concurrent tasks, then the time for executing any remainder is estimated. Then, it is added to the original time estimate $T(numTasks, c)$.
    \item If there is a tie and several constraints result in the same execution time for a given number of ready tasks, then the highest constraint is used because it result in the minimum I/O congestion. 
\end{itemize}

It should be noted that the objective function is reevaluated and a new constraint is set -if necessary- every time new execution requests of an auto-constrained task arrive to the scheduler. This approach allows tuning the value of the constraint depending on the number of I/O auto-constrained tasks.

\section{Evaluation}
\label{sec:evaluation}

In this section, we show the improvement that I/O aware PyCOMPSs can achieve in the total performance of applications with different workloads. 

We start this section with describing the infrastructure of the MareNostrum 4 supercomputer and its storage architecture (Section \ref{subsec:infrastructure}). Next, we present a brief description of the applications used in the evaluation, their I/O workload characteristics and their performance results (Section \ref{subsec:exps}). Finally, the section ends with presenting the experiments results that show the impact of the hyperparameters of the auto constraints on applications performance (Section \ref{sec:hyper}).

\subsection{Infrastructure}
\label{subsec:infrastructure}

~\

The MareNostrum 4 supercomputer \cite{marenostrum} of the \\Barcelona Supercomputeing Center (BSC) is composed of 3,456 nodes. Each node has two Intel Xeon Platinum chips, each with 24 processors for a total of 48 cores per node. The MareNostrum 4 supercomputer contains two types of nodes: low memory and high memory. The low memory nodes contain 92 GB main memory whereas the high memory nodes contain 370 GB of main memory.  

\begin{figure}[htbp]
    \centering
    \includegraphics[width=\linewidth]{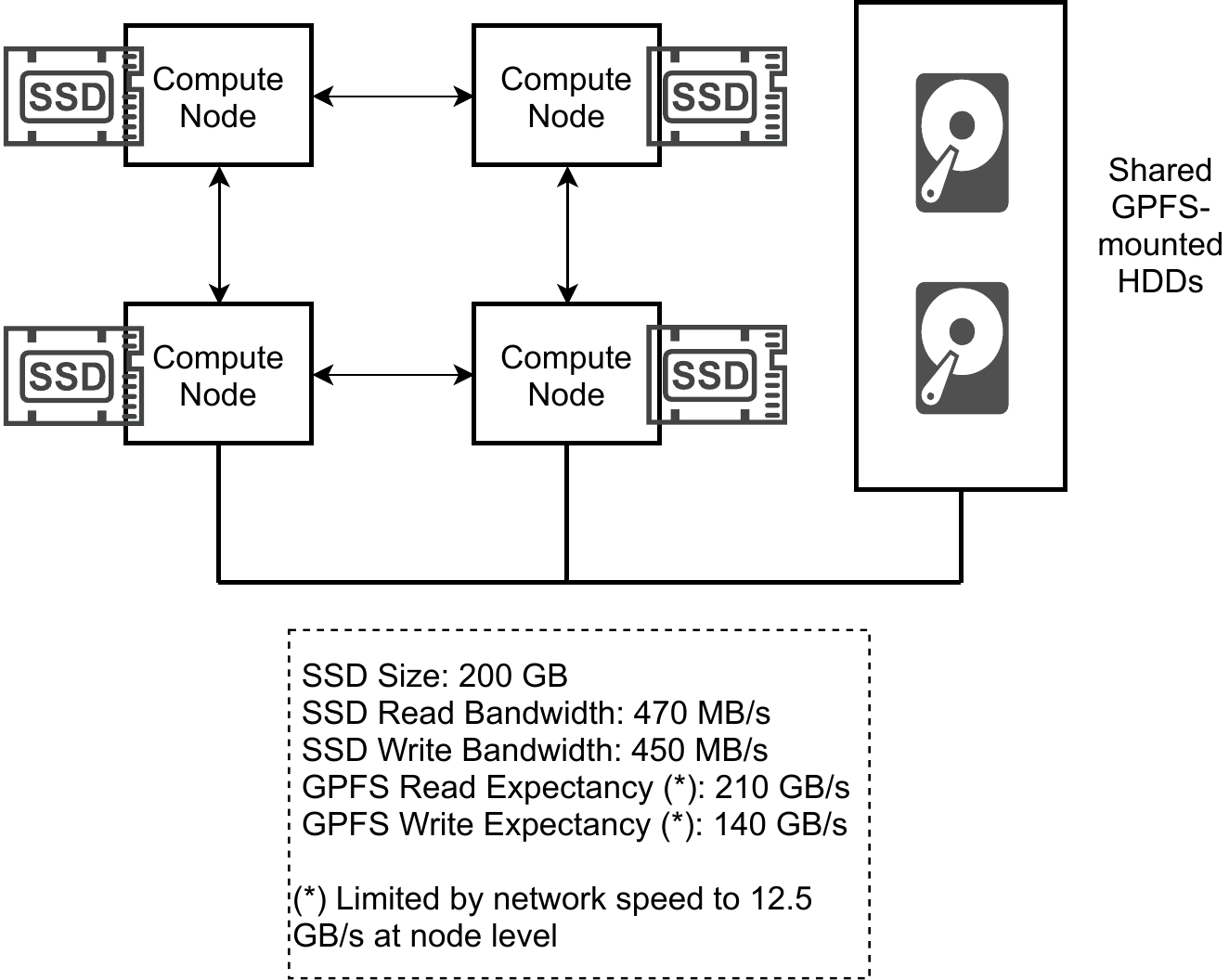}
    \caption{High-Level Overview of the Storage Infrastructure on the MareNostrum 4 Supercomputer}%
    \label{fig:mn}%
\end{figure}

Figure \ref{fig:mn} shows a high-level overview of the MareNostrum 4 supercomputer storage infrastructure. All nodes have access to a shared Hard Disk Drive (HDD) with total capacity of 14 PetaBytes mounted with the IBM General Parallel File System (GPFS). Moreover, each node has a local Solid State Drive (SSD) with a capacity of 200 GB and bandwidth of 470 MB/s and 450 MB/s for reading and writing respectively. 

In all our experiments, the GPFS is used to store the input data and final results -if any- of the applications. The GPFS has a performance expectancy of 210 and 140 GB/s for read and write operations respectively. This performance is limited by the network speed at the node level to 12.5 GB/s. As the GPFS servers are shared between all the users of the MareNostrum4 supercomputer, its performance in practice is much lower than the maximum theoretical values. 

In addition to that, the GPFS is also used in our experiments as a shared working directory to store tasks logs and dependencies between tasks. Therefore, no node-to-node data transfer is required. 

Node-local SSD disks are used as \textit{Burst Buffers} to checkpoint the intermediate results of the applications. On the one hand, as they are used exclusively by the nodes running the experiments, they offer better performance than the globally shared HDD-backed GPFS. This is because the whole bandwidth of the SSDs are dedicated for our experiments and no interference occurs from the experiments of other MareNostrum4 users. On the other hand, they offer a controlled environment in which performance benefit can be planned and expected. Using SSDs as a caching layer to absorb intensive I/O of applications has been discussed in previous I/O research \cite{burstbuffer,warp,nersc}.

\subsection{Experiments}
\label{subsec:exps}

~\

We implemented three different I/O intensive applications with PyCOMPSs. Each application exhibits a different I/O workload which allows for evaluating the impact of the I/O awareness capabilities in different scenarios. These three applications are: 
\begin{itemize}
    \item The \textit{HMMER} application: an application that produces \textit{Homogeneous I/O workload}; there is only one task that execute I/O in the application. In addition to that, for a given input size, the checkpointing task writes the same amount of I/O every time it is called during the lifetime of the application. This application is intended to show the impact of using I/O awareness capabilities on I/O throughput, I/O task time and total time. In addition to the behaviour of auto-tunable constraints with homogeneous I/O workloads. 

    \item The \textit{Variants Discovery Pipeline}: exhibits a \textit{Heterogeneous I/O workload}, because it has more than one checkpointing task. Each checkpointing task writes different amount of data to the disk. This application is intended to show the behaviour of auto-tunable constraints when different I/O workloads.

    \item \textit{Kmeans}: an iterative algorithm to test the effect of the number of available tasks on the total performance of the application when using auto-tunable constraints.

\end{itemize}

In all the experiments, the I/O non-aware PyCOMPSs implementation (i.e. no I/O tasks nor storage bandwidth constraints) is used as the baseline version. Moreover, for the \textit{HMMER} and \textit{Variants Discovery Pipeline}, we launched several runs of the I/O aware implementation. Each run has a different setting of the storage bandwidth constraint for the I/O tasks. These runs include: 
\begin{itemize}
    \item A non-constrained run where I/O tasks are used to execute I/O workload but no storage bandwidth constraints are used.
    \item Several runs with increasing values of static storage bandwidth constraint.
    \item Two runs with the both types of the auto-tunable constraints. For each of the runs with an auto-tunable constraint, we show graphs of its learning phase progress during application execution.
\end{itemize}

It should be noted that for the \textit{HMMER} application and the \textit{Variants Discovery Pipeline}, reading I/O tasks have been used in order to read input data. However, they do not offer any performance benefit because they do not overlap with compute tasks.

To test the effectiveness of our proposals for solving the problem of I/O congestion, writing I/O tasks in all experiments is avoided using system buffers by flushing the data to storage devices. This is achieved by using the \textit{fsync} call of the os library of the Python programming language {\cite{pythonos}}.

All the experiments were run on 12 high-memory \\MareNostrum 4 nodes plus one node dedicated as the master node. The master node runs the master component of the COMPSs runtime and manages the execution without taking part in any computation.

\subsubsection{HMMER Application}
\label{subsubsec:homo}

~\

\begin{figure}[htbp]
    \centering
    \includegraphics[width=\linewidth]{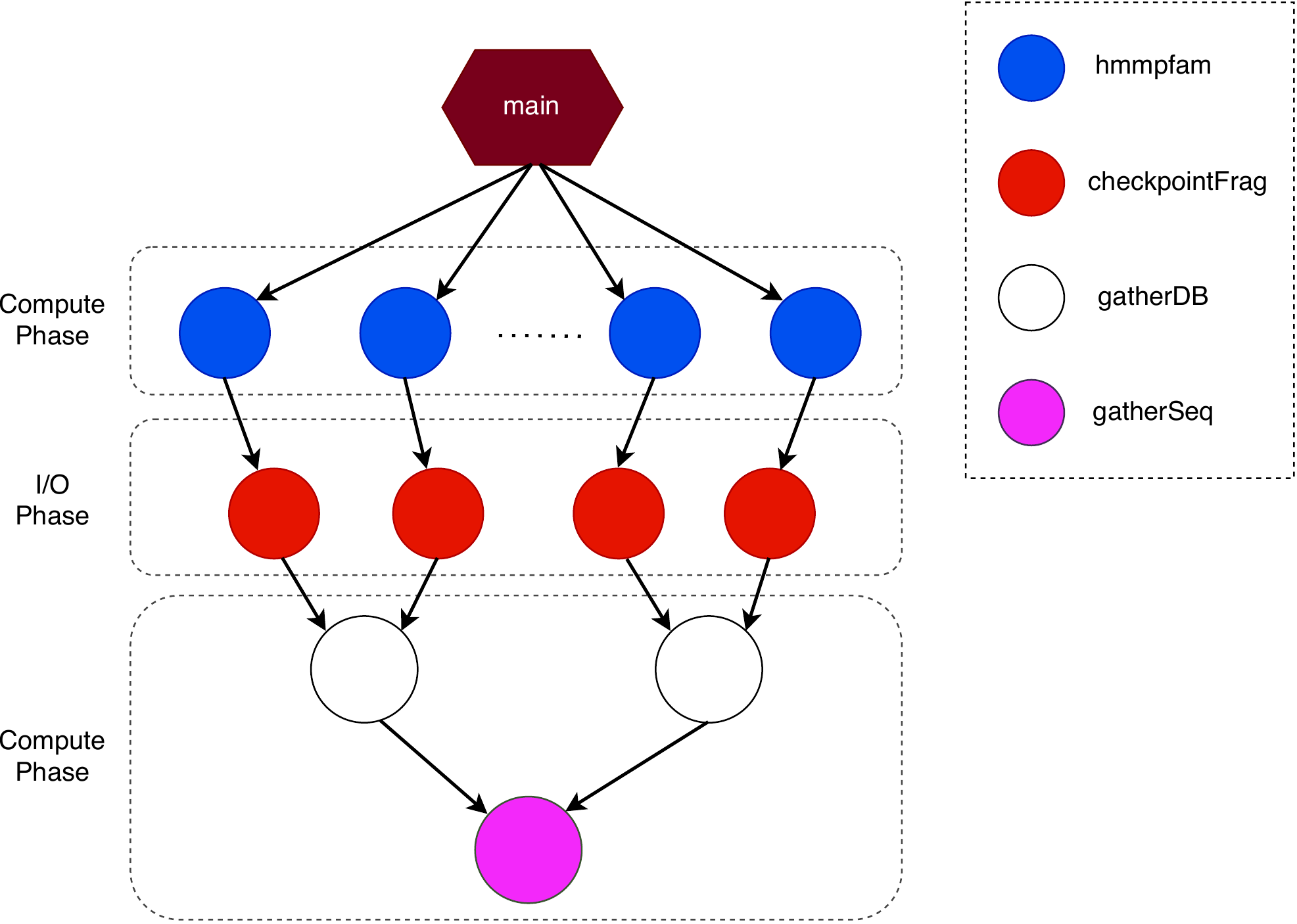}
    \caption{Task Skeleton of the HMMER application}%
    \label{fig:hmmer}%
\end{figure}

The \textit{HMMER} Application is used for searching sequence databases for sequence homologous proteins or nucleotide sequences using a variant of Hidden Markov Models (HMM) called \textit{profile-HMM}. It takes two inputs: a sequence database and a sequence file. Figure \ref{fig:hmmer} depicts a sample PyCOMPSs tasks dependency graph of the application for a small data-set. This application first splits the sequence file and sequence database into multiple fragments. A \textit{hmmpfam} task is called for each sequence and database fragment. The \\ \textit{hmmpfam} task calls the HMMER tool \cite{hmmer} on its sequence fragment and database fragment. Each \textit{hmmpfam} task has as a \textit{checkpointFrag} successor task that is responsible for checking the results of the HMMER tool. Later, the application calls a \textit{gatherDB} task which gathers the results obtained of running a single sequence fragment against all database fragments. Finally, all the sequence fragments are gathered into one single file in the \textit{gatherSeq} task.

For running the experiments of this application, we used as inputs a 64.5 GB HMM protein-families database (pfam) and a sequence file that contains 14,942,208 sequences with a total size of 3.2 GB. Databases and sequence files are available on the ftp servers of the European Bioinformatics Institute (EMBL-EBI) \cite{ebi} which hosts up-to-date sequences, databases and software widely used by academics and life science researchers. 


We set the number of database fragments and the sequence fragments to 48 each. This means that every run of the application will have 2,304 \textit{hmmpfam} tasks followed by the same number of \textit{checkpointFrag} tasks. Each \textit{checkpointFrag} task writes 290 MB of data to a separate file on the node-local SSD disk. As each of the 12 worker nodes used in this experiment has 48 cores, then the maximum number of tasks that can run at the same time across the whole system is 576 tasks. Consequently, the application will be executed in multiple compute-IO phases. 

\begin{figure}[htbp]
    \centering
    \includegraphics[width=\linewidth]{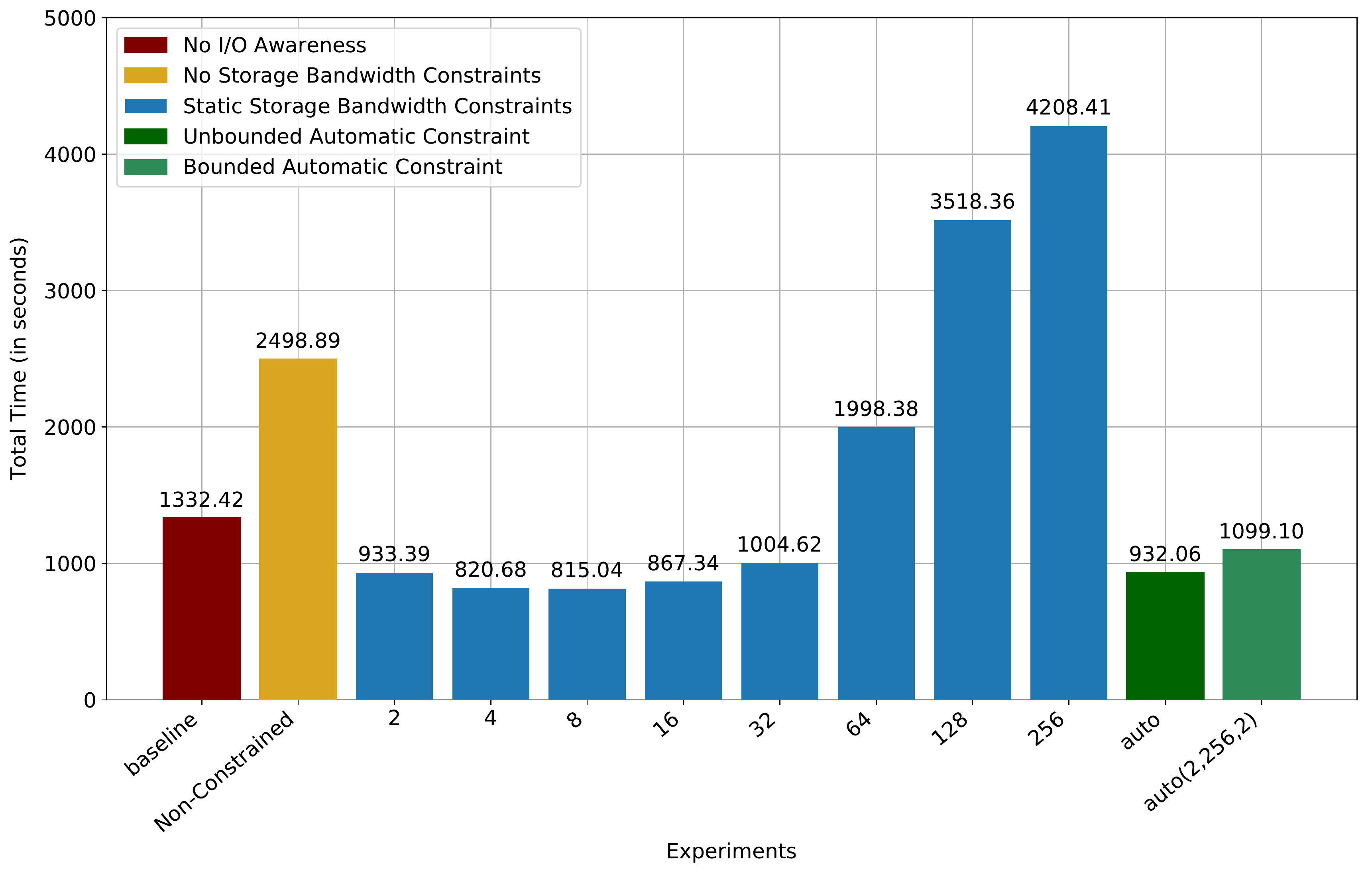}
    \caption{Experimental results for the HMMER Application}%
    \label{fig:homo_perf}%
\end{figure}

Figure \ref{fig:homo_perf} presents the performance results of the application. In Figure \ref{fig:homo_perf}, the red bar represents the baseline run where non of the I/O capabilities (i.e. I/O tasks and storage bandwidth constraint) are used. Whereas the yellow bar represents a non-constrained run where only one capability of I/O aware PyCOMPSs is used; declaring the \textit{checkpointFrag} as an I/O task but without using any storage bandwidth constraint for the I/O tasks. The blue bars represent runs with both I/O capabilities of PyCOMPSs, each run using a higher static storage bandwidth constraints. Finally, there are two bars: one represents using the unbounded automatic storage bandwidth constraint and the other represents a run with a bounded auto storage bandwidth constraint of \textit{auto(2,256,2)}. For the non-constrained run we used 500 I/O executors, whereas 225 I/O executors are used for the static and auto storage bandwidth runs.

As can be noted in Figure \ref{fig:homo_perf} using the I/O awareness capabilities of PyCOMPSs can achieve more than 38\% performance improvement in the total time of the application compared the baseline version. In the baseline run, the \textit{checkpointFrag} are treated as compute tasks, no overlap between I/O and computations occurs and only 48 \textit{checkpointFrag} task can be executed at a time. However, using I/O tasks without setting any storage bandwidth constraint (as in the non-constrained run) results in a much worse total time than the baseline. Even though the execution of the \textit{checkpointFrag} I/O tasks is overlapped with the execution of the \textit{hmmpfam} compute tasks, the effect of the I/O congestion negatively affects the total time of the application, since no constraints are used.

Nevertheless, continuing with Figure \ref{fig:homo_perf}, as we start setting a storage bandwidth constraint for the I/O tasks, application's total time starts to decrease not only because I/O and compute tasks overlap execution but also I/O congestion is controlled. As the value of the storage bandwidth constraint increases, the total time of the application improves until a certain point where it starts to deteriorate again. Indeed, increasing the value of the constraint decreases the maximum number of concurrent I/O tasks. Even though executing less I/O tasks concurrently minimizes I/O congestion and improves I/O task time, this improvement in I/O task time does not compensate the decreased task parallelism. This is most apparent when using a storage bandwidth constraint of 256 where only one I/O task is allowed to run at a time. In this case, even though the whole I/O bandwidth is entirely dedicated for the currently running I/O task, the sequential execution of I/O tasks drastically harms the total time of the application. 

Furthermore, it can be observed in Figure \ref{fig:homo_perf} that both runs with the auto storage bandwidth constraint achieve total time improvements compared to the baseline experiment. However, it can be noted that the total time when using a bounded auto constraint is worse than the total time when using the unbounded auto constraint. 

\begin{figure}[htbp]
    \centering
    \includegraphics[width=\linewidth]{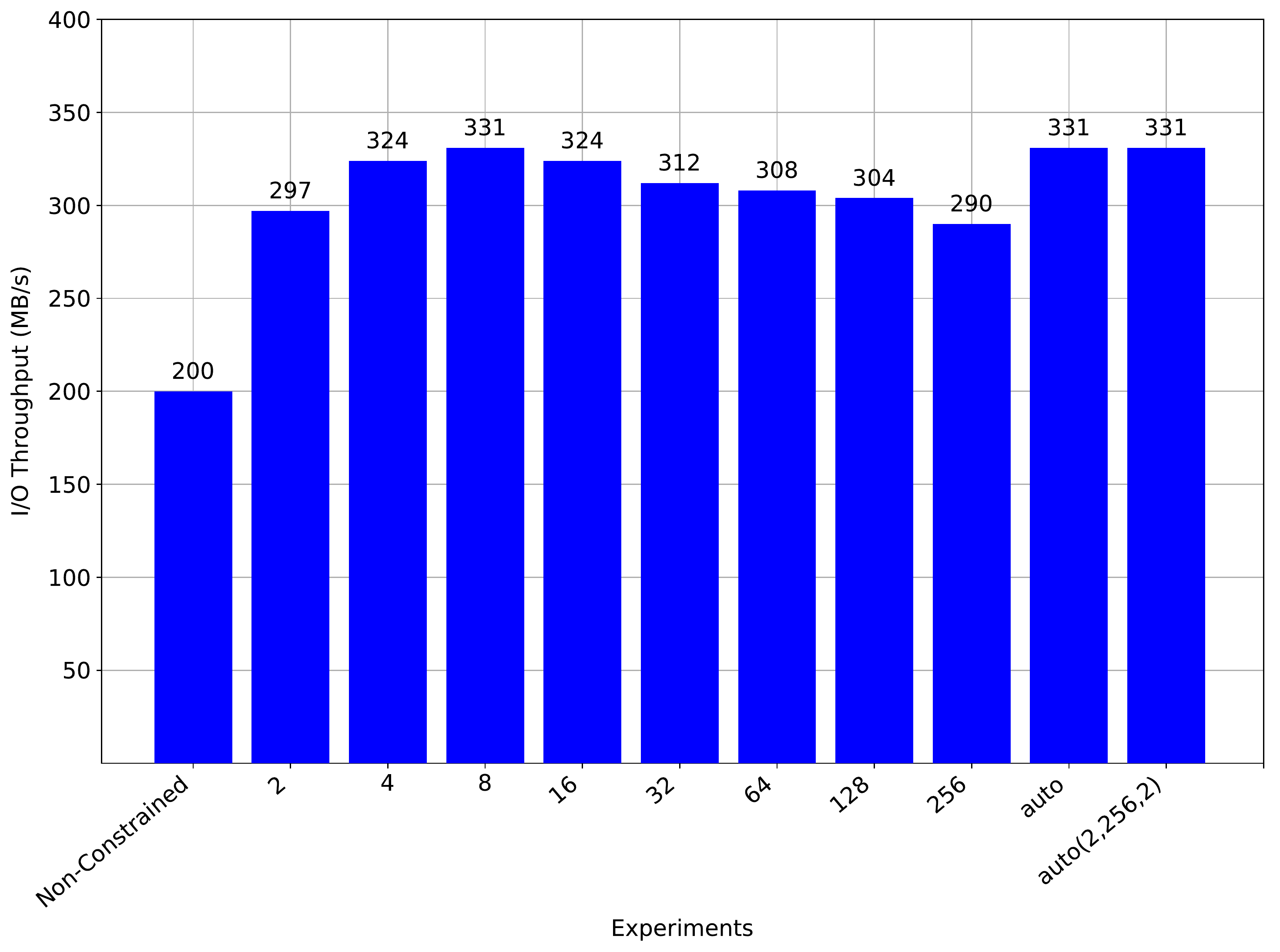}
    \caption{Achieved I/O Throughput in the HMMER Application}%
    \label{fig:throughput}%
\end{figure}

\begin{figure}[ht]
    \centering
        \subfigure[Unbounded Auto-tunable Constraint]{
        \includegraphics[width=0.8\linewidth]{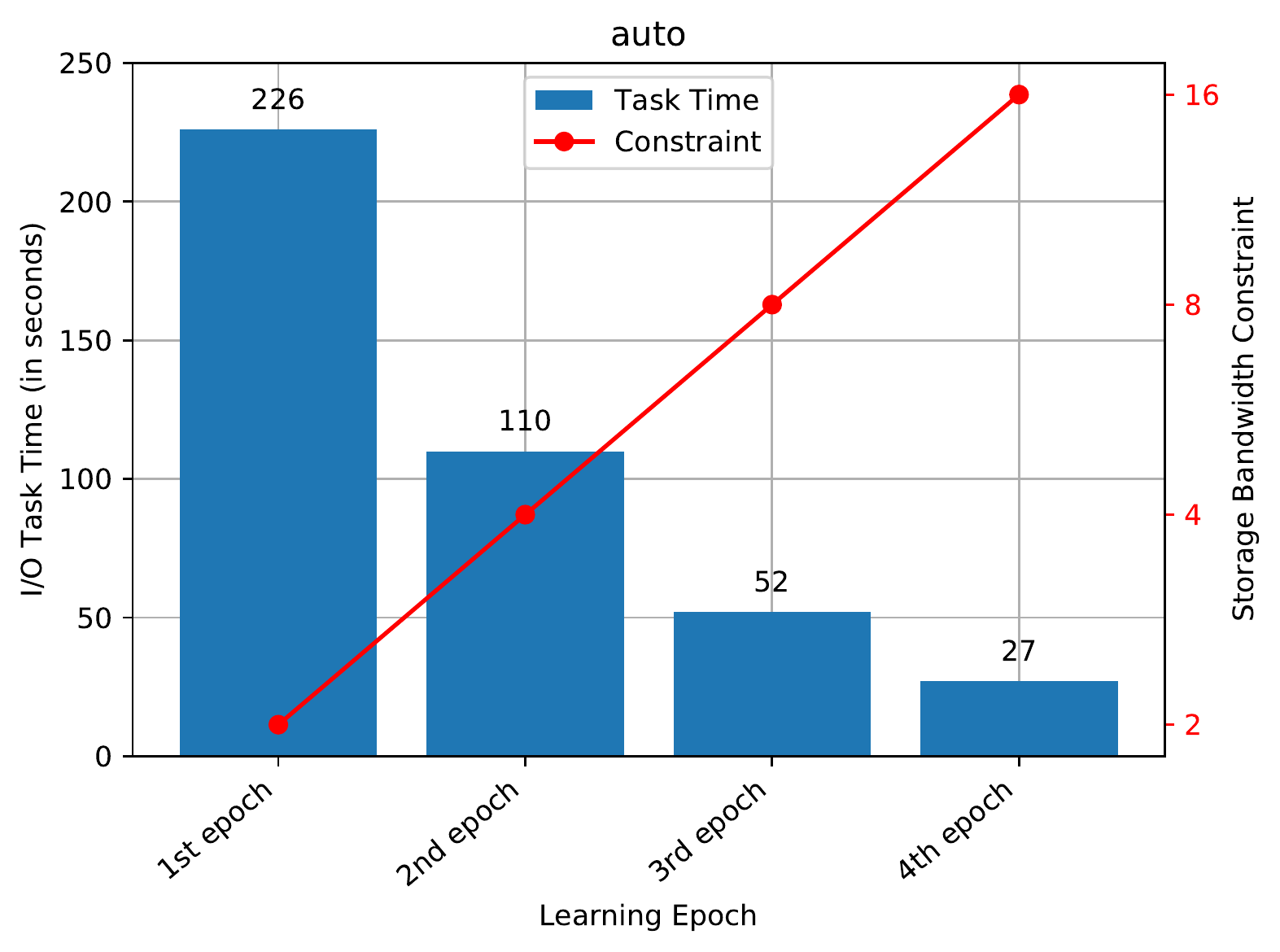}
        \label{fig:homo_auto}
    }
    \subfigure[Bounded Auto-tunable Constraint]{
        \includegraphics[width=0.8\linewidth]{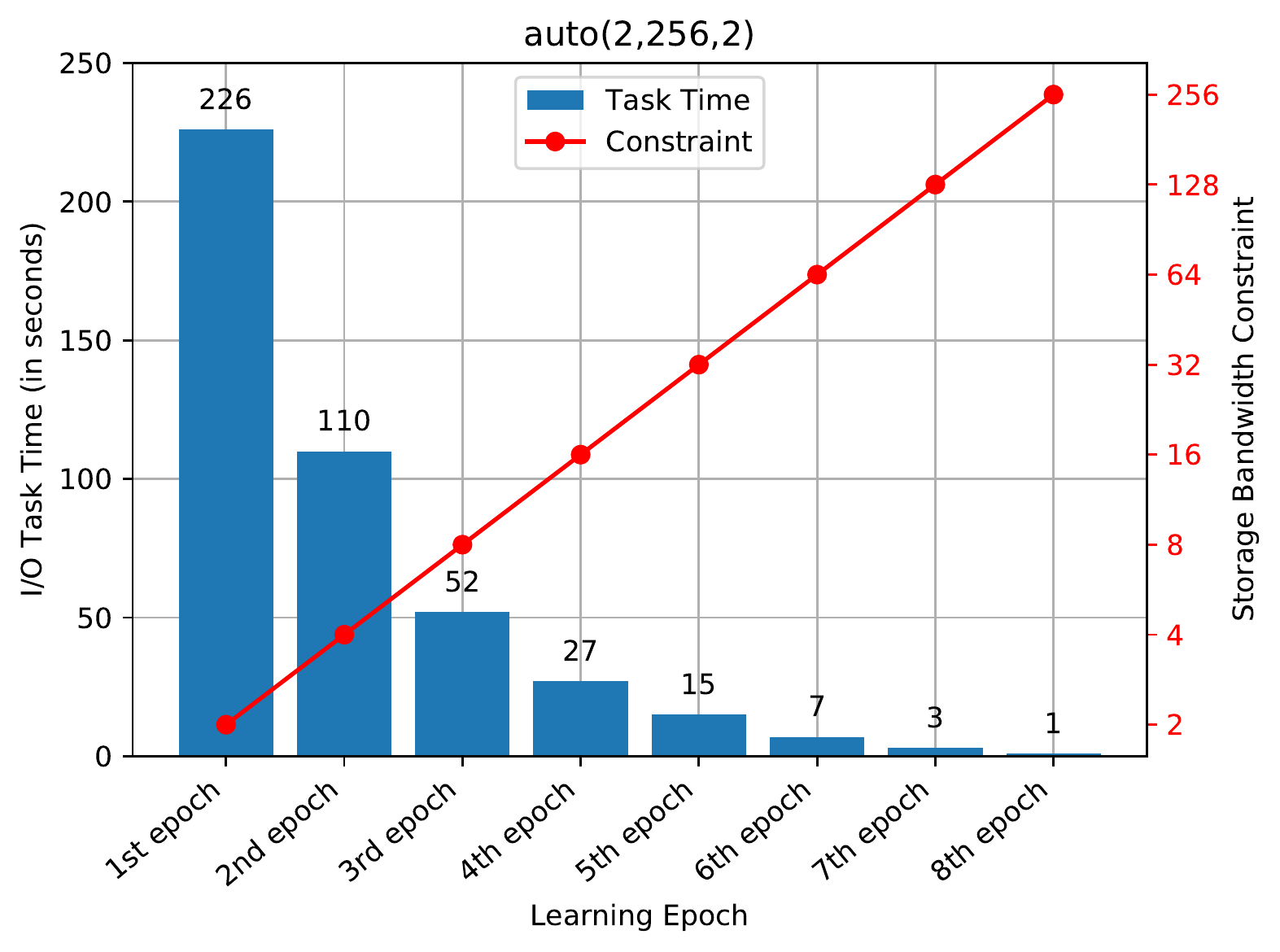}
        \label{fig:homo_autominmax}
    }
    \caption{Auto-tunable Constraints Learning Phase Progress in the HMMER Application}
    \label{fig:hmmer_autos}
\end{figure}

Figure {\ref{fig:throughput}} presents the achieved I/O throughput for I/O tasks. The non-constrained experiment has the worst I/O \\ throughput due to the increased and uncontrolled I/O congestion. As we start to control the number of I/O tasks running concurrently by using bandwidth constraints, I/O congestion decreases and the achieved I/O throughput begins to increase until it reaches the peak value when a constraint of 8 is used (which is the same value at which the application has the best total time in Figure {\ref{fig:homo_auto}}). As the constraint value keeps increasing, the number of I/O tasks running concurrently decreases, therefore I/O throughput slightly decreases because the local-SSDs of the nodes are not fully utilized. Furthermore, it can be observed that both of the auto constraints achieve peak I/O throughput similar to using the optimal constraint.  

In order to understand the auto constraints behaviour, we refer to Figure \ref{fig:hmmer_autos} that shows the progress of the learning phases of both auto constraints during application's execution time. Figure \ref{fig:homo_auto} depicts the progress of the learning phase when using an unbounded auto constraint. First, the runtime sets the initial constraint to 2, because this run used 225 I/O executors on each worker node to handle the execution of I/O tasks. After the end of the first learning epoch the runtime registers the average I/O task time during this epoch and doubles the value of the constraint to progress the learning phase. When the second epoch ends, in order to decide whether to continue or abort the learning phase, the runtime checks whether the I/O task time in the second learning epoch is at least half of the I/O task time in the previous epoch. Since this condition is met, the runtime registers the average I/O task time during the second epoch. Next, the runtime progresses the learning phase until it stops after the fourth epoch because the continuation condition is violated; the task time in the fourth epoch is not at least half of the task time in the third epoch. Upon the termination of the learning phase, the runtime applies the auto constraint registry, that now contains the average I/O task time in three learning epochs, to objective function \ref{eq:fn}. Finally, the runtime sets the constraint to 8 which is the value that minimizes the execution time of the auto constrained \textit{checkpointFrag} I/O tasks ready for scheduling.


Similarly, the learning phase progress of the bounded auto constraint \textit{auto(2,256,2)} is depicted in Figure \ref{fig:homo_autominmax}. Using this type of auto constraints, the runtime starts the first epoch with the minimum constraint value provided in the \verb|@constraint| decorator in the user code: 2. After the end of the first epoch, the runtime registers the average I/O task time and progress the learning phase by multiplying the current constraint value by the value of delta provided in the user code: 2. Therefore, the second learning epoch has a constraint of 4. The learning phase keeps progressing in this manner until the current value of the constraint becomes bigger than the maximum value provided by the user: 256. Therefore, the learning phase stops after the eighth learning epoch. Now that the runtime has filled the auto constraint registry, it uses the auto constraint registry to minimize the execution which in this case is 8. 

As the bounded auto constraint spends more time in the learning phase, its application total time is worse than the unbounded auto constraint that follows a stricter and shorter learning process. In addition to that, the bounded auto constraint has a more fine-grained \textit{auto constraint registry}, since it tries higher constraint values. During most of the execution time, the final constraint value of the bounded auto constraint and the unbounded auto constraint is the same (in this application, this constraint value is 8). However, for a certain number of scheduling-ready \textit{checkpointFrag} I/O tasks, the fine-grained auto constraint registry of the bounded auto constraint may result in a different constraint value than the unbounded auto constraint for smaller number of tasks. However, the value of the constraint is re-adjusted and the minimization function is re-evaluated every time a new \textit{checkpointFrag} I/O task arrives to the scheduler and the number of scheduling-ready \textit{checkpointFrag} I/O tasks increases. Therefore, if the runtime sets a high constraint value for a certain number of scheduling-ready tasks, this constraint \\value will be adjusted in the next scheduling iteration.



\subsubsection{Variants Discovery Pipeline}
\label{subsec:hetero}

~\

\begin{figure}[!htpb]
    \centering
    \includegraphics[width=\linewidth]{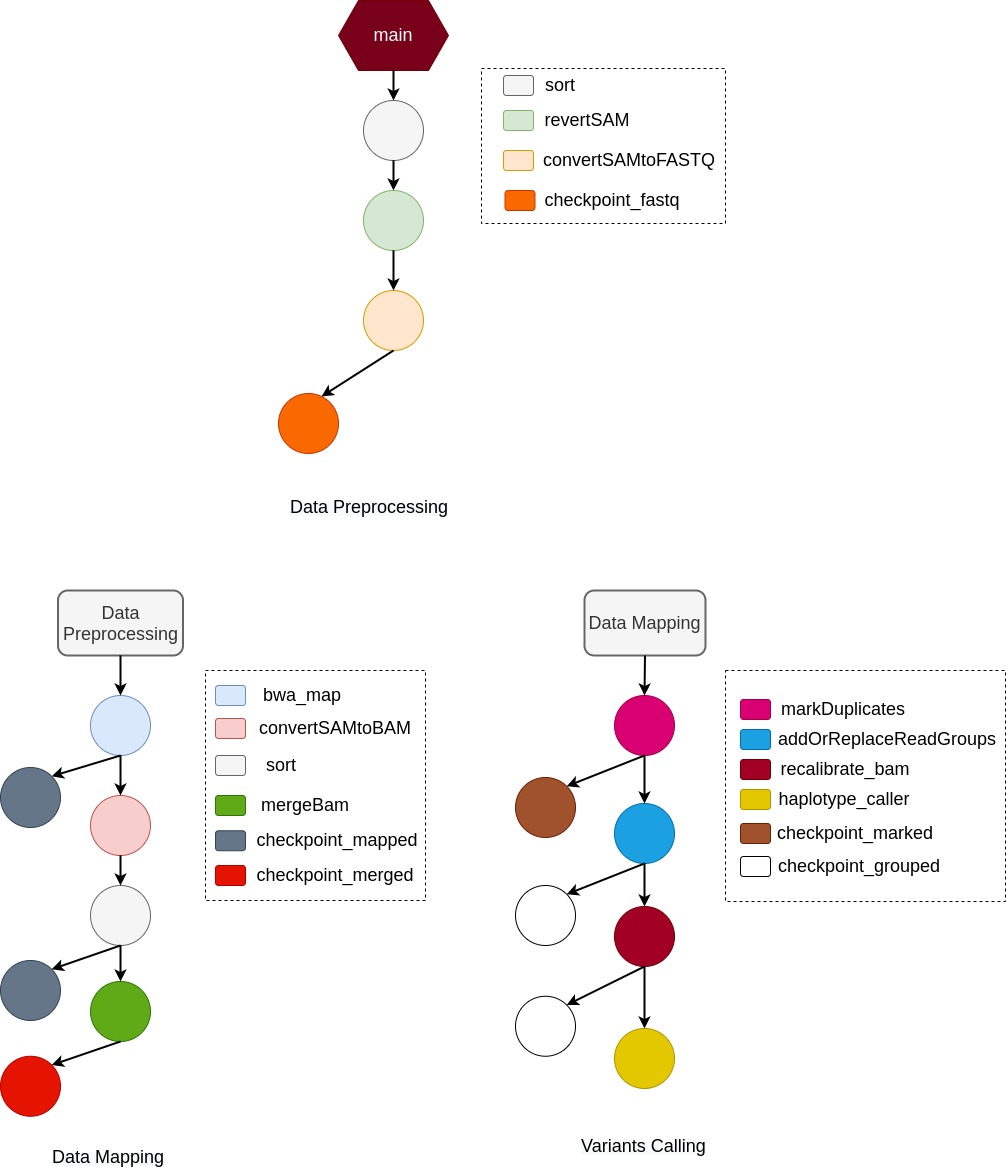}
    \caption{Task Skeleton of the Variants Discovery Pipeline}%
    \label{fig:bioinf}%
\end{figure}

The \textit{Variants Discovery Pipeline} is popular in the field of bioinformatics and computational genomics. The purpose of this pipeline is to discover genomic variants in sequence data. Figure \ref{fig:bioinf} illustrates the PyCOMPSs tasks dependency graph of this pipeline. Since the pipeline performs a lot of operations, we split it for into three phases for visualization purposes: Data Preprocessing, Data Mapping and Variant Calling. We defined a checkpointing task for each major step in the pipeline to checkpoint the results of the pipeline so far. We define a major step in the pipeline as the last step in each processing phase (e.g. \textit{convertSAMtoFASTQ} at the end of \textit{Data Processing} phase) or any step that is not easily recomputed (e.g. after \textit{bwa\_map} in the \textit{Data Mapping} phase). 

When two compute tasks produce output of the same size, they call the same checkpointing task. For instance, the \textit{bwa\_map} that maps its input to the reference genome and the \textit{sort} task that sorts the mapped sequence, use the \textit{checkpoint\_mapped} task because they produce almost the same size of output data.   

We launched the experiments for this application with 1,728 sample sequence files. Each sequence file has a size of 72 MB in a compressed gzipped format. For each input, the application launches a separate pipeline to discover its variants. The input sequence files and the meta-data are publicly available on the GATK Broad Institute servers \cite{gatk} which is a well-known resource for providing human sequencing data (e.g. sample sequences, genome references, variants databases, etc.). Table \ref{table:sizes} lists the checkpointing tasks in the application and the data sizes that each task writes.

\begin{table}
 \centering
 \caption{Amount of data written by checkpointing tasks}
 \begin{tabular}[t]{||c c||} 
 \hline
 Task & Amount of Data Produced \\ [0.5ex] 
 \hline\hline
 checkpoint\_fastq & ~162 MB  \\ 
 \hline
 checkpoint\_mapped & ~290 MB \\
 \hline
 checkpoint\_merged & ~330 MB  \\
 \hline
 checkpoint\_marked & ~596 MB \\
 \hline
 checkpoint\_grouped & ~615 MB \\ [1ex] 
 \hline
 \end{tabular}
\label{table:sizes}
\end{table}

\begin{figure}[htbp]
    \centering
    \includegraphics[width=\linewidth]{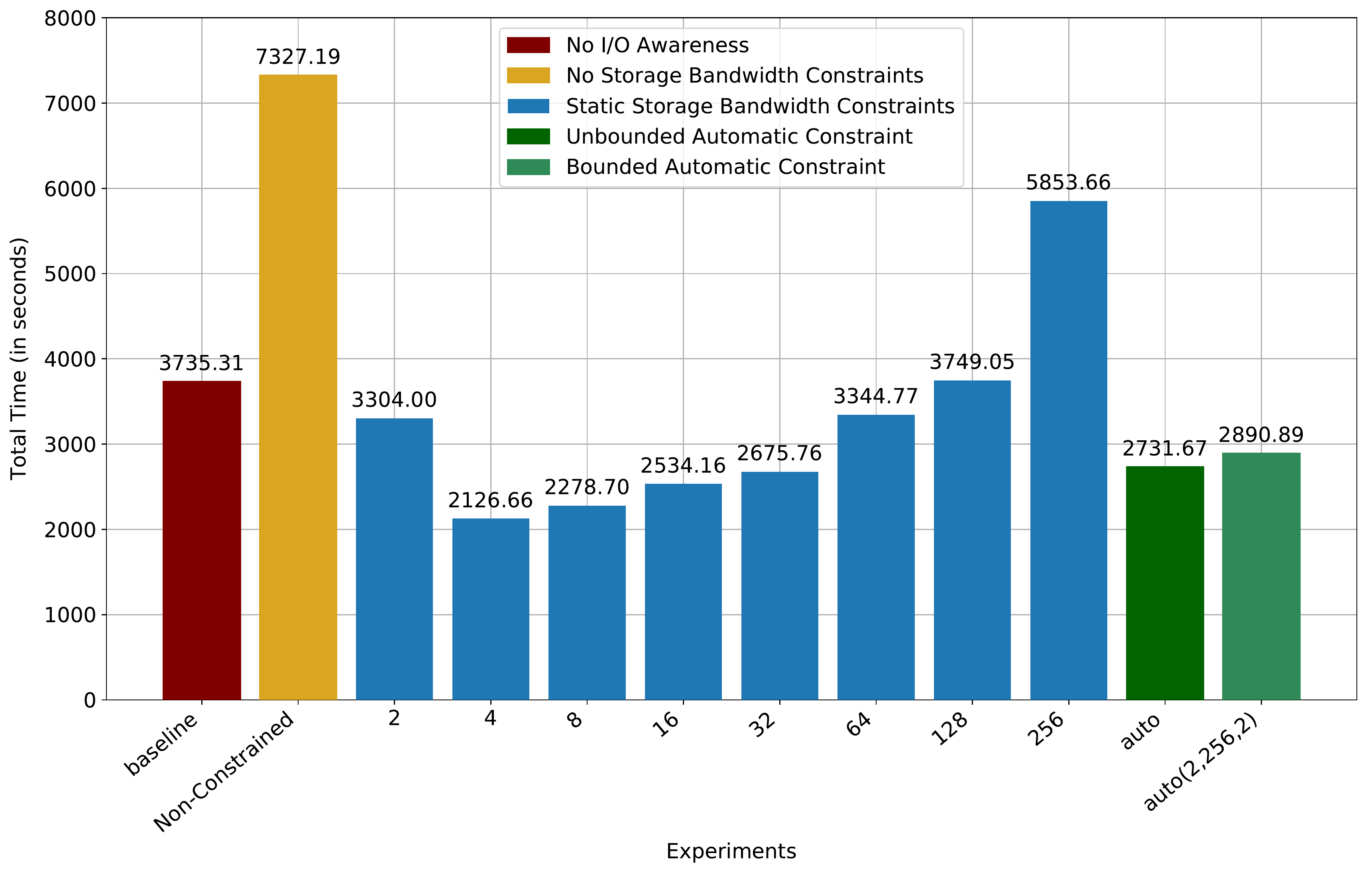}
    \caption{Experiment Results of Variant Discovery Pipeline}%
    \label{fig:hetero_perf}%
\end{figure}

Figure \ref{fig:hetero_perf} presents the performance results of different runs of the application. Using both capabilities of I/O awareness (i.e. I/O tasks and storage bandwidth constraints) can achieve up to 43\% performance improvement in the total time of the application compared to the baseline run. The non-constrained run has the worst total time because of the I/O congestion that happens as all the I/O tasks concurrently access the node-local SSD disk of the same worker node. In this run, a maximum of 325 I/O tasks are allowed to run concurrently because 325 I/O executors are used. Using the storage bandwidth constraint immediately mitigates the I/O congestion problem and the total time starts to improve. However, as the static storage bandwidth constraint increases, the total time starts to degrade due to the decreased level of task parallelism. Moreover, both auto constraints runs achieve performance improvement comparable to the optimal total time when using a static constraint of 4 with some overhead incurred due to the time spent in the learning phase. 

\begin{figure}[!htbp]
    \centering
    \includegraphics[width=0.8\linewidth]{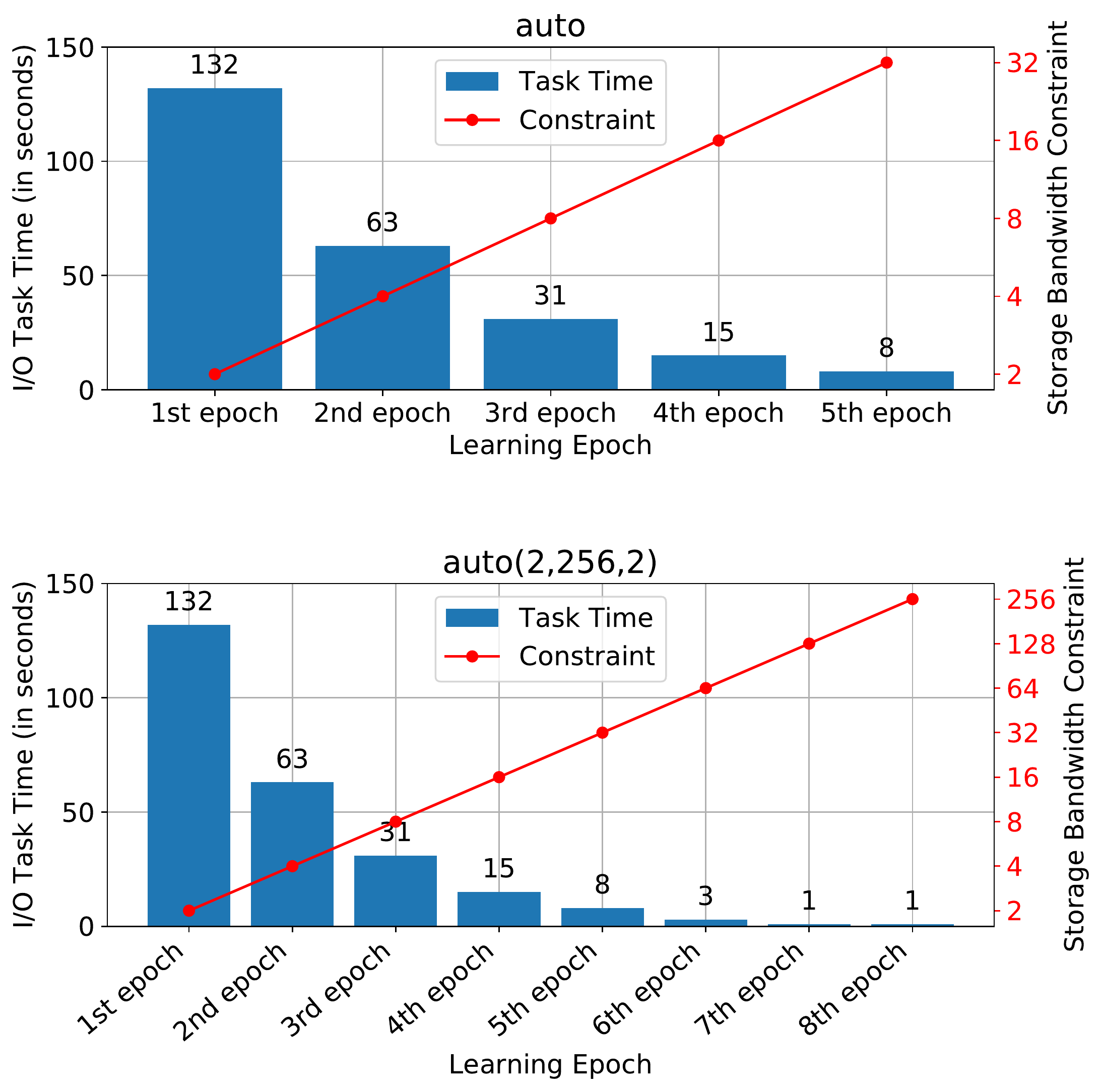}
    \caption{Learning Phase of \textit{checkpoint\_fastq} Task.}%
    \label{hetero:1}%
\end{figure}

\begin{figure}[!htpb]
    \centering
    \includegraphics[width=0.8\linewidth]{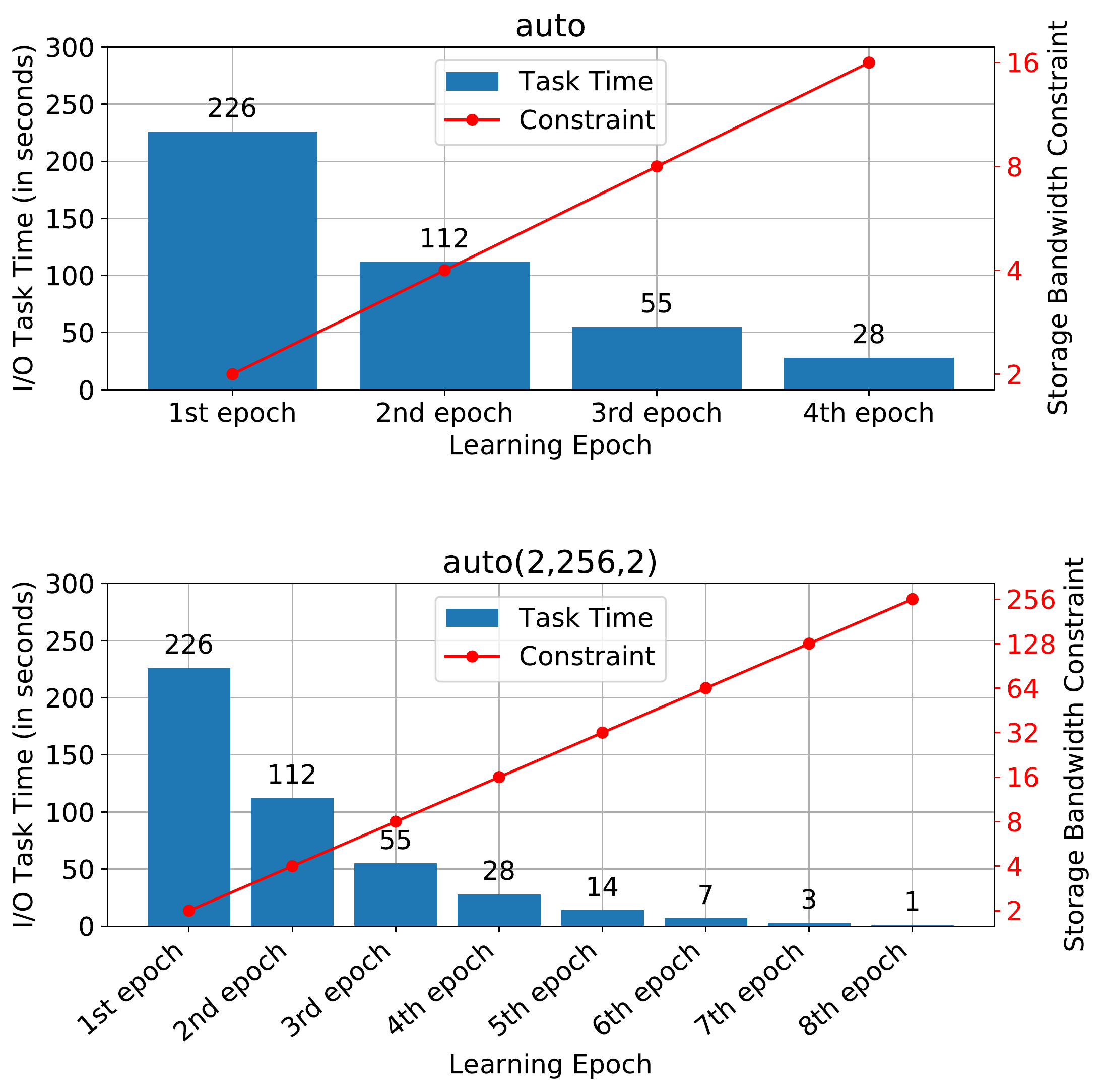}
    \caption{Learning Phase of \textit{checkpoint\_mapped} Task.}%
    \label{hetero:2}%
\end{figure}

\begin{figure}[htbp]
    \centering
    \includegraphics[width=0.8\linewidth]{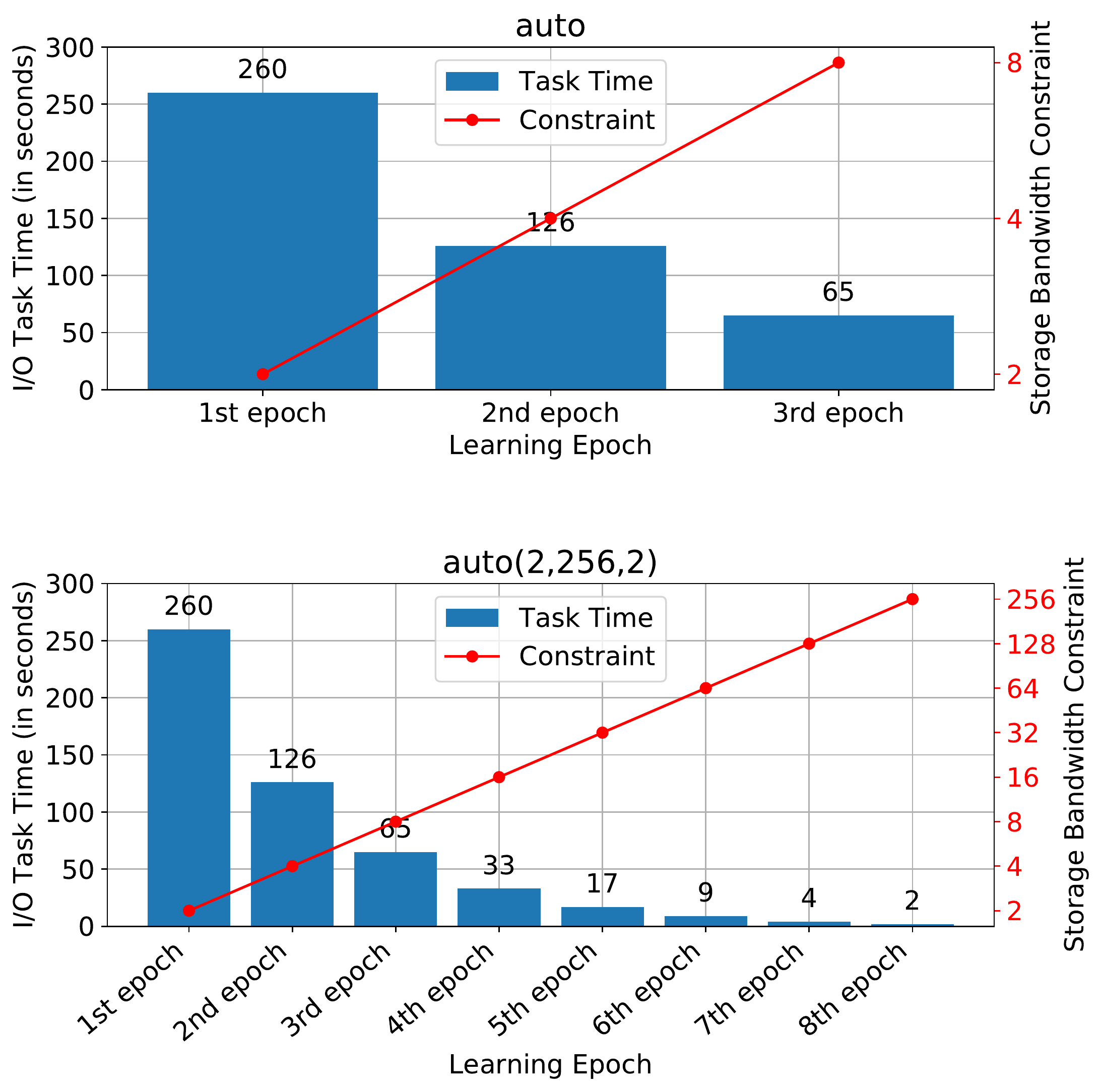}
    \caption{Learning Phase of \textit{checkpoint\_marked} Task.}%
    \label{hetero:3}%
\end{figure}

\begin{figure}[htbp]
    \centering
    \includegraphics[width=0.8\linewidth]{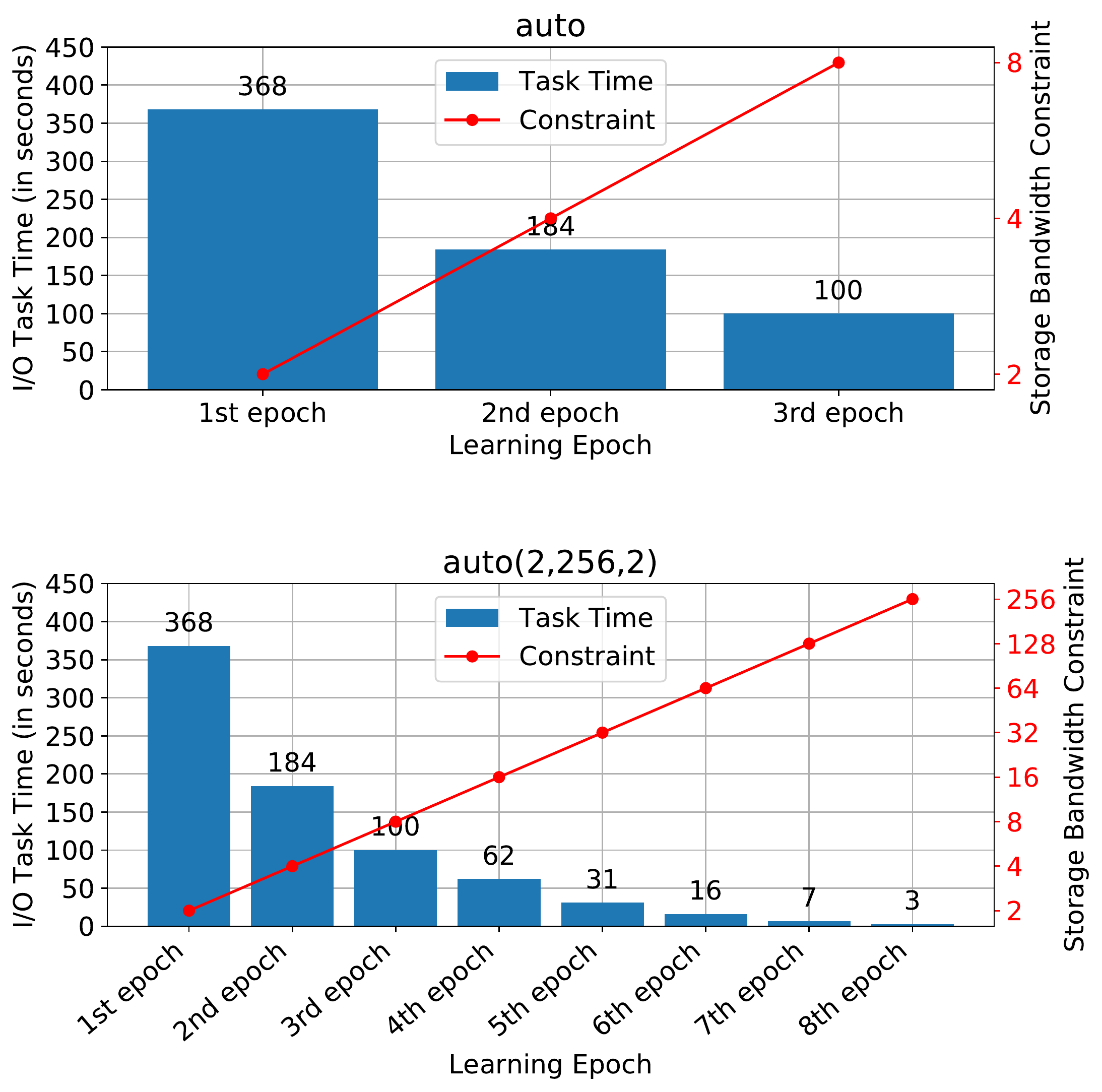}
    \caption{Learning Phase of \textit{checkpoint\_merged} Task.}%
    \label{hetero:4}%
\end{figure}

\begin{figure}[htbp]
    \centering
    \includegraphics[width=0.8\linewidth]{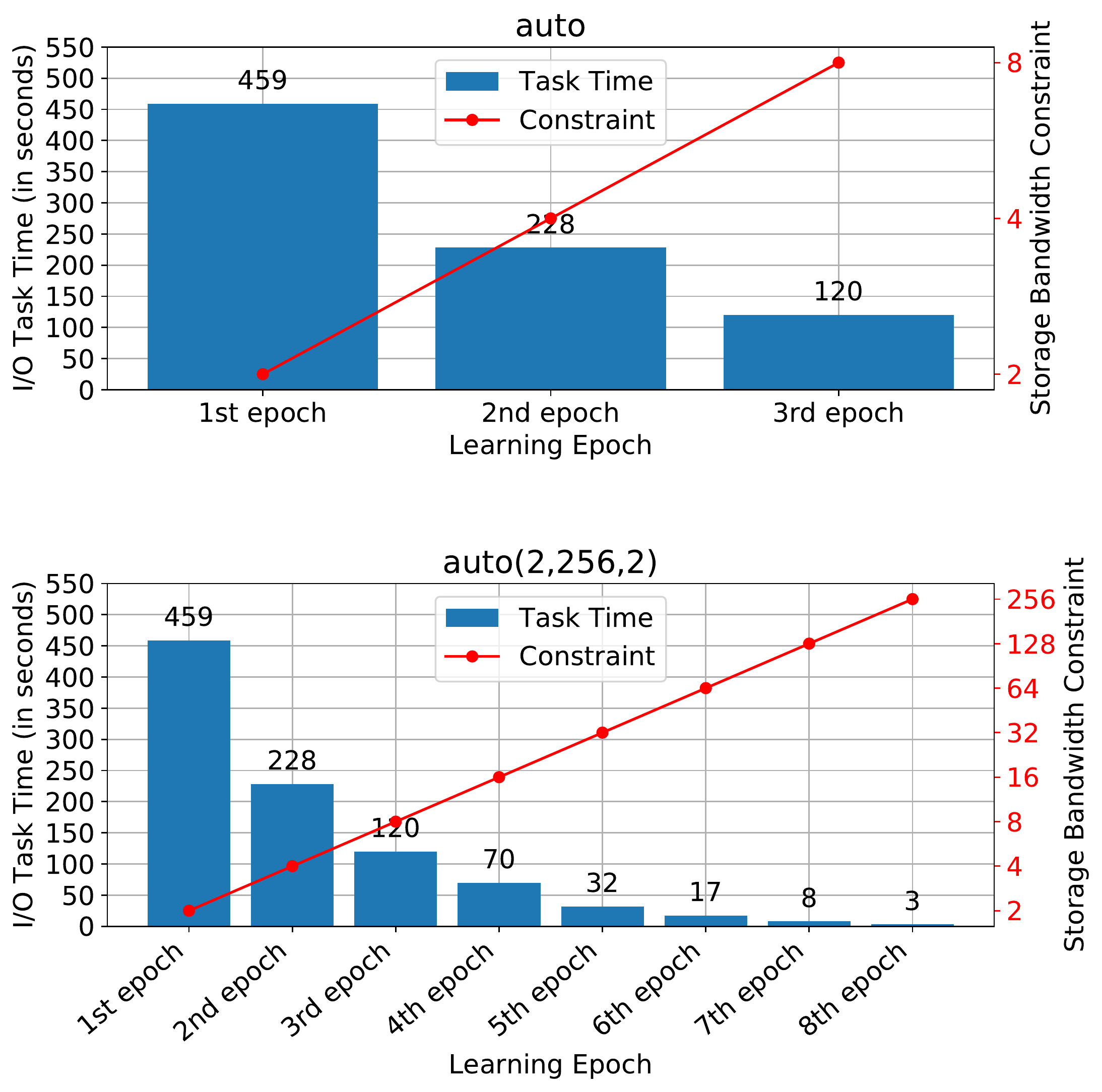}
    \caption{Learning Phase of \textit{checkpoint\_grouped} Task.}%
    \label{hetero:5}%
\end{figure}

It should be noted that in the static constraint runs, the same static constraint is used for all the checkpointing tasks mentioned in Table \ref{table:sizes}.  However, in the auto constraint runs, the final value of the auto constraint is different for each checkpointing task. Each of these checkpointing tasks has its own learning phase, and the objective function is evaluated for each of them separately. \Crefrange{hetero:1}{hetero:5} show the learning phase progress for each checkpointing task with the unbounded and bounded auto constraint.  

It can be noticed from \Crefrange{hetero:1}{hetero:5} that each checkpointing task goes through its own learning phase. The runtime uses the constraint value that will optimize the execution of the I/O workload of this task independently of the other checkpointing tasks. Table \ref{table:final_constraints} lists the final auto constraints that were used for each checkpointing I/O task. 

\begin{table}[!htpb]
 \centering
 \caption{Constraint values for the checkpointing Tasks}
 \begin{tabular}[t]{||c c||} 
 \hline
 Task & Constraint \\ [0.5ex] 
 \hline\hline
 checkpoint\_fastq & 16  \\ 
 \hline
 checkpoint\_mapped & 8 \\
 \hline
 checkpoint\_merged & 4  \\
 \hline
 checkpoint\_marked & 4 \\
 \hline
 checkpoint\_grouped & 4 \\ [1ex] 
 \hline
 \end{tabular}
\label{table:final_constraints}
\end{table}

Although different constraints are used for each checkpointing I/O task, using auto constraints can achieve a total time performance close to the optimal total time achieved when static constraint of 4 is used. This is possible because the runtime sets the auto constraint for the \textit{checkpoint\_merged} and \textit{checkpoint\_grouped} tasks to 4, which is the constraint that leads to the best total time. Therefore, unlike static constraints where a certain constraint value maybe optimal for one checkpointing task but not optimal for the others, auto constraints will use the constraint that achieve best possible I/O task time and total execution time.

Note that the constraint choice of both: \textit{checkpoint\_mer-\\ged} and \textit{checkpoint\_grouped} has bigger impact on the total time than the other checkpointing tasks because these two tasks are executed at the end of the pipeline where there are no compute tasks that can hide the effect of using a bad constraint.



\subsubsection{Kmeans Application}
\label{subsec:iterative}

~\

In order to evaluate the impact of the number of I/O tasks on the time of the learning phase of the auto constraints and consequently the application total time, we run multiple experiments with the Kmeans application as an example of iterative applications. The \textit{Kmeans Application} is a well-known machine learning algorithm that is widely used for different purposes such as cluster analysis in data mining fields. The Kmeans application follows an iterative process where it groups a set of multidimensional points into a number of clusters following a nearest mean distance rule. By changing the number of iterations, we change the number of tasks that will be executed in the application. Increasing the number of iterations will generate more tasks to be executed, whereas decreasing the number of iterations will decrease the number of tasks available for execution.

The dataset considered for evaluating the Kmeans application is composed of 10,000,000 points of 1000 dimensions, 3,000 centers and 500 fragments. Each of the checkpointing tasks writes 109 MB to the SSD storage disks.

\begin{figure}[htbp]
    \centering
    \includegraphics[width=\linewidth]{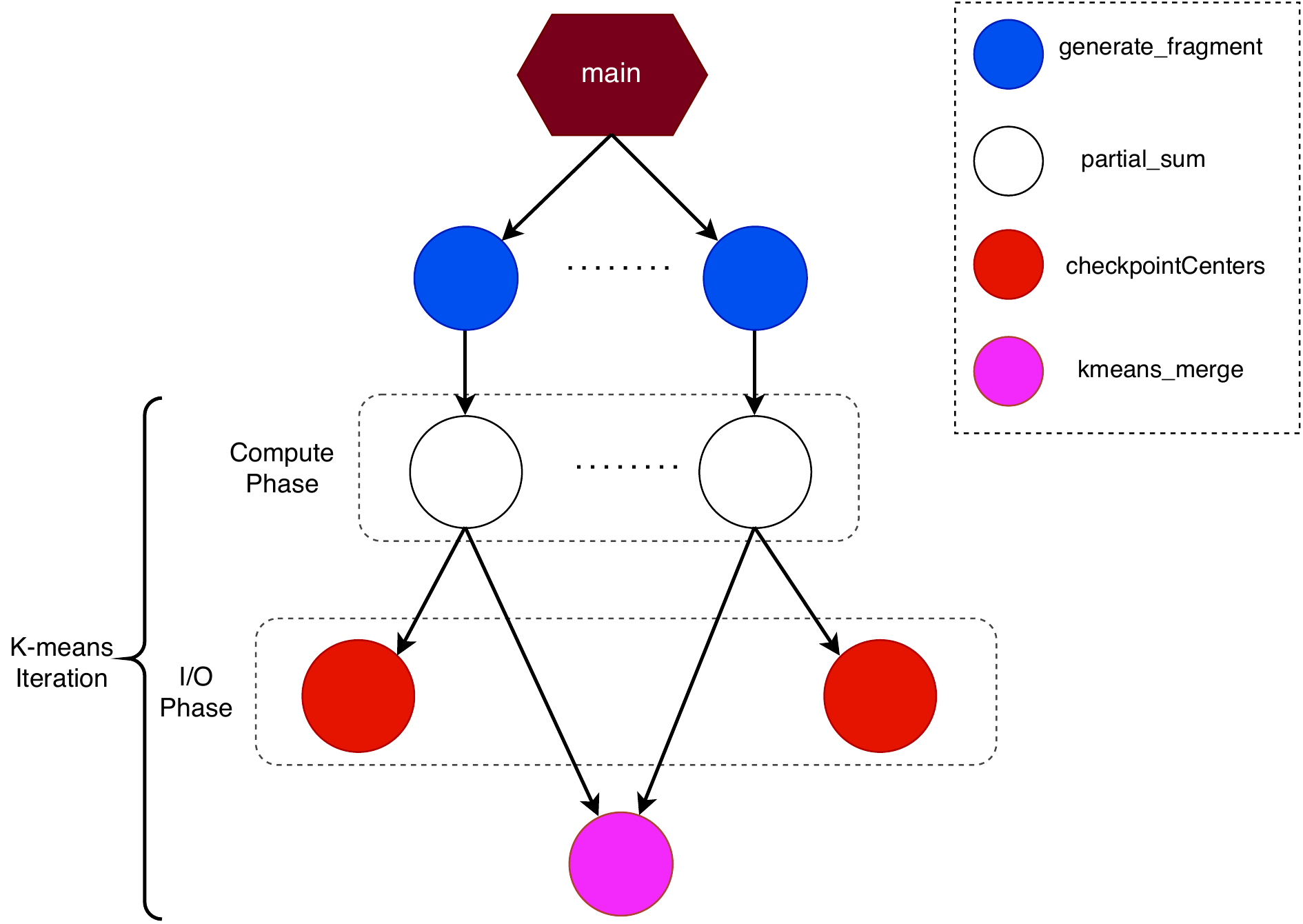}
    \caption{Task Skeleton of the Kmeans Application}%
    \label{fig:kmeans}%
\end{figure}

Figure \ref{fig:kmeans} shows a sample  task dependency graph of the Kmeans application implemented with PyCOMPSs for a sm-\\all dataset.  A \textit{generate\_fragment} generates fragments of random data given a specific seed. In each iteration, the \textit{partial\_sum} task is called on each fragment to calculate the distance of each point to all cluster centers. The new centers are checkpointed using the \textit{checkpointCenters} task.

\begin{figure}[htbp]
    \centering
    \includegraphics[width=\linewidth]{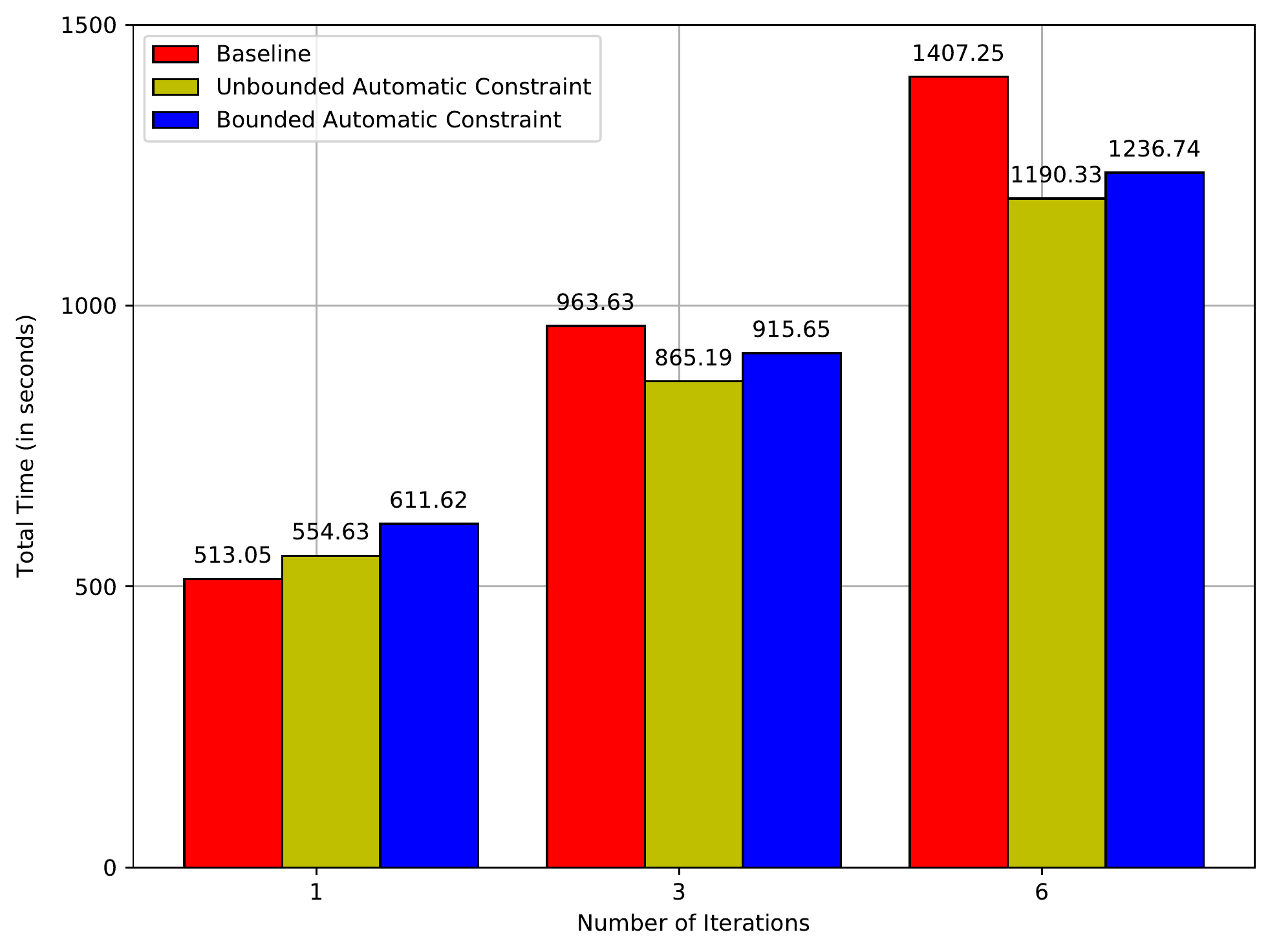}
    \caption{Kmeans Application with different number of iterations}%
    \label{fig:kmeans_itrs}%
\end{figure}

Figure \ref{fig:kmeans_itrs} shows the experiments results of the Kmeans applications with different number of iterations. It can be noticed that with a single iteration the results of both of the auto constraints experiments do not show a performance improvement. This can be explained due to the small number of auto-constrained checkpointing tasks in the application. Out of 500 checkpointing tasks to be executed, the unbounded auto constraint uses 435 checkpointing tasks for learning, whereas the bounded auto constraint uses 446 checkpointing tasks. Therefore, after the learning phase ends, a very small number of checkpointing tasks remains to take advantage of the results of the learning phase.

In order to validate this conclusion, we repeated the experiments but this time with more number of  iterations (3 and 6). In the case of 3 iterations, the total number of checkpointing task available for execution increases to 1500 tasks. Consequently, the number of auto-constrained tasks available for execution increases and we start getting performance improvement for both of the auto constraints. This gain increases with increasing the number of iterations because the application can make up the time spent in the learning phase and more I/O tasks overlap with the execution of compute tasks.

\subsection{Hyperparameters Experiments}
\label{sec:hyper}

~\

In this section, we evaluate the performance impact of changing the values of the hyperparameters of both of the auto constraints. In the case of the bounded auto constraint, these parameters are \textit{min}, \textit{max} and \textit{delta} set by the user in the \verb|@constraint| decorator. Whereas in the case of the unbounded auto constraint, the hyperparameter is the number of I/O executor threads per worker node. 

\begin{figure}[!htbp]
    \centering
        \subfigure[Experiments of the HMMER Application]{
        \includegraphics[width=\linewidth]{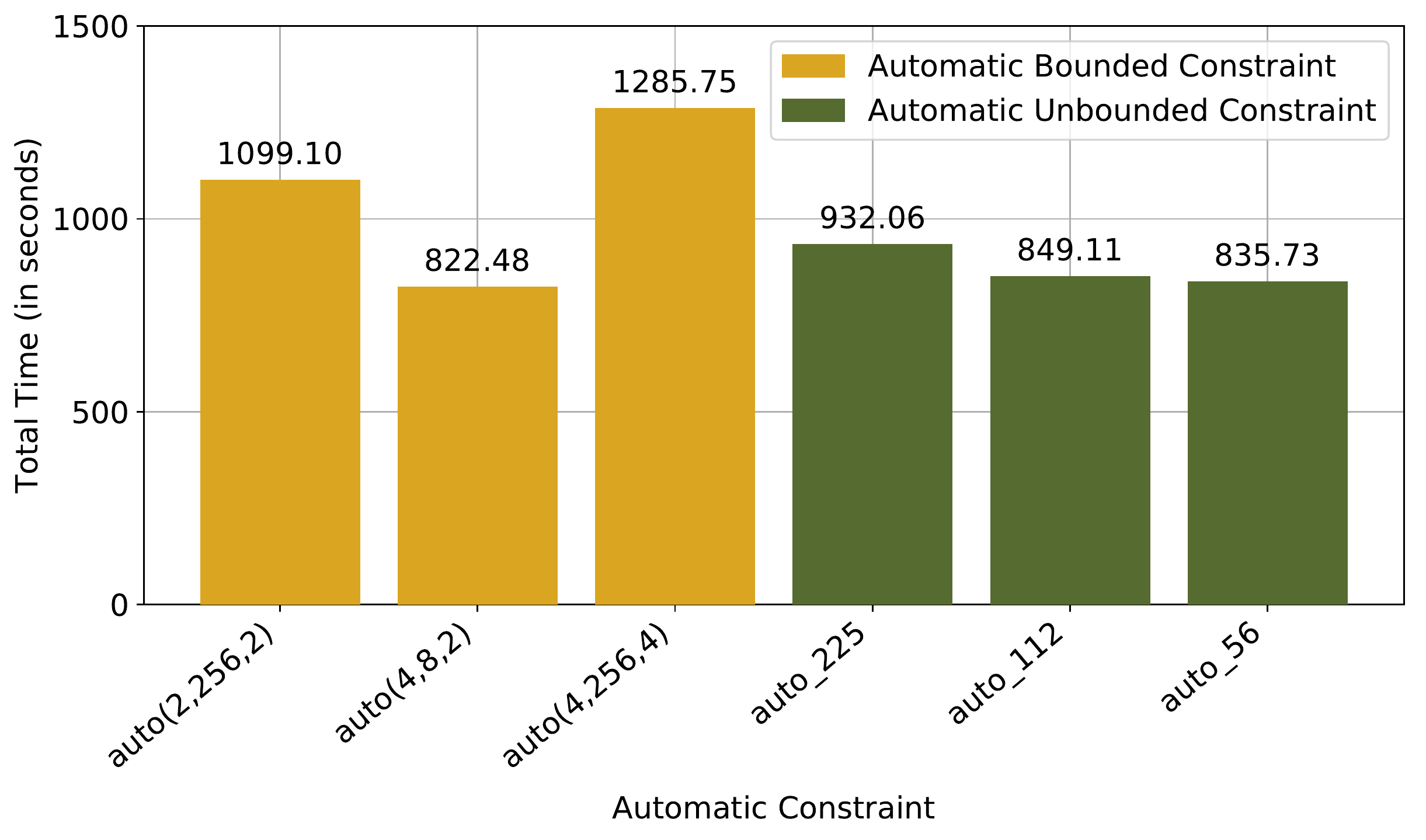}
        \label{fig:hyper_homo}
    }
    \subfigure[Experiments of the Variants Discovery Pipeline]{
        \includegraphics[width=\linewidth]{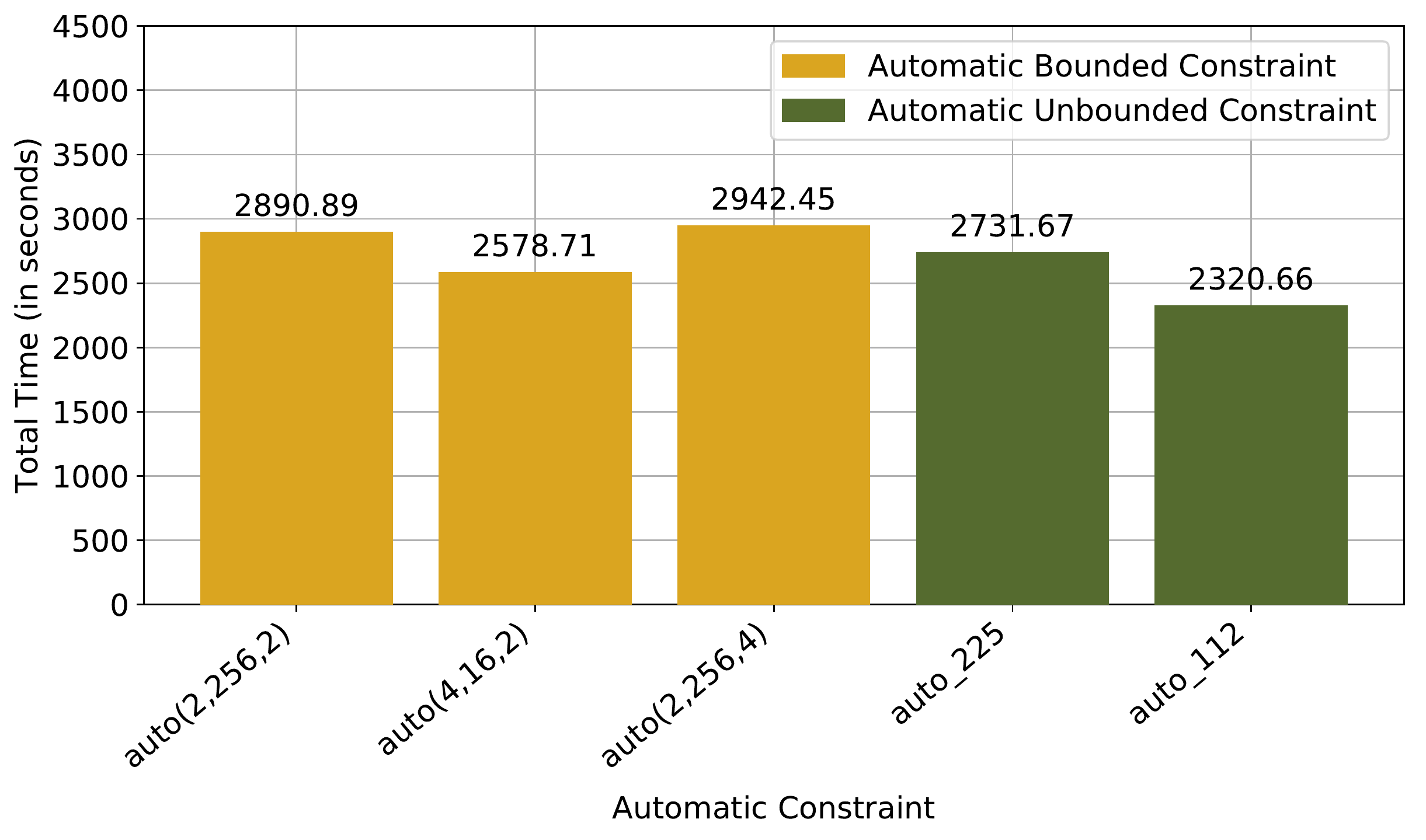}
        \label{fig:hyper_hetero}
    }
    \caption{Hyperparameters Experiments}
    \label{fig:hyper}
\end{figure}

To this end, we repeated the experiments of the HMMER application (homogeneous I/O workload) and the Variants Discovery Pipeline (heterogeneous I/O workload) using auto constraints with different hyperparameter values. Figure \ref{fig:hyper} shows the experiments results for both applications. It can be noted that in both applications, changing the values of the hyperparameters impacts the total performance. In the HMMER application (Figure \ref{fig:hyper_homo}), the optimal constraint is 8 so setting the constraint to \textit{auto(2,256,2)} incurs a long learning phase. Whereas adjusting the \textit{min} and \textit{max} values to \textit{auto(4,16,2)} shortens the learning phase and results in a better total time. In another run, Setting a big value of \textit{delta} like in the constraint \textit{auto(4,256,4)} to speed up the learning phase resulted in a worse total time because a big value of \textit{delta} skipped the optimal constraint (i.e. 8). 

Furthermore, continuing with Figure \ref{fig:hyper_homo}, we can see that the unbounded auto constraint achieves better results. Using an unbounded auto constraint with 225 I/O executors incurs a longer learning time because the starting constraint is 2. However, the learning time in this case is not as long as \textit{auto(2,256,2)} due to the strict conditions of the unbounded auto constraint. Decreasing the number of I/O executors results in a better total time since it approaches the optimal constraint: 8. With 112 I/O executors, the constraint of the first learning epoch will be set to 4 whereas it will be set to 8 with 56 I/O executors.

Likewise, a similar behaviour can be seen with the Variants Discovery Pipeline (Figure \ref{fig:hyper_hetero}). Adjusting the boundaries of the hyperparameters like in \textit{auto(4,16,2)} to decrease the learning phase improves the total time. Also, using a larger value of delta (\textit{auto(2,256,4)}) may result in an increase in the total time because optimal constraints are skipped. On the other hand, using unbounded constraints achieve better total time; using 225 I/O executor achieves better total time than \textit{auto(2,256,2)} because of the shorter learning phase. Moreover, using 112 I/O executors achieves better total time because the learning epoch starts with constraint value 4 which is the optimal constraint for the two checkpointing tasks at the end of the pipeline (i.e. \textit{checkpoint\_merged} and \textit{checkpoint\_grouped}) and the learning phase is shorter.

\section{Conclusions and Future Work}
\label{sec:conclusion}

This paper proposes to enable I/O awareness in Task-based programming models. I/O awareness increases the amount of parallelism inherent in I/O intensive applications by taking advantage of the optimization opportunities possible due to I/O workloads and compute-I/O patterns. With an I/O aware task-based programming model, opportunities for compute-I/O execution overlap can be exploited. In addition to that, I/O performance bottlenecks such as I/O congestion can be mitigated, thus resulting in total time performance improvements. 

In order to take advantage of the I/O awareness capabilities, it is necessary to separate I/O from computation when programming applications. Hence, an I/O aware runtime system can take advantage of I/O properties to improve the performance of applications.
 


We implemented the I/O awareness capabilities in the PyCOMPSs tasking framework and evaluated it with different I/O workloads. The evaluation demonstrated that significant total performance improvement can be achieved compared to the default I/O non-aware PyCOMPSs implementation. 


As future work, we plan to extend our proposals to address the case when shared resources are used to absorb the I/O of applications. In this scenario, certain assumptions and modified mechanisms have to be adopted to take into account the I/O performance variability on shared resources. Furthermore, we aim to extend the auto-tunable constraints to support the inference of other constraints such as as memory size and number of processes in MPI tasks executions.

\section{Acknowledgements}

This work is partially supported by the European Union through the Horizon 2020 research and innovation programme under contracts 721865 (EXPERTISE Project) by the Spanish Government (TIN2015-65316-P) and the Generalitat de Catalunya (contract 2014-SGR-1051).

\bibliographystyle{unsrtnat}

\bibliography{cas-refs}

\bio{pics/toma}
Hatem Elshazly is a marie-curie Ph.D. student in the Computer Architecture department at the Technical University of Catalonia (DAC-UPC). Since 2017, he has been working as a research engineer at the Barcelona Supercomputing Center (BSC) optimizing task-based programming models for I/O and memory critical applications. He holds a MSc. degree in the optimization of data intensive applications on distributed infrastructures from Nile University, Egypt (2016). He is currently collaborating in the EXPERTISE multidisciplinary European project under the Horizon 2020 research and innovation programme targeting the optimization of I/O intensive applications. He also collaborated in a multidisciplinary international project with the Harvard Medical School targeting the performance and cost optimization of personalized medicine workflows for clinical use. His current research interests include parallel programming models, high performance computing, mitigating the I/O and memory bottlenecks and the management of heterogeneous distributed infrastructure.
\endbio

\bio{pics/jorgee}
Jorge Ejarque holds PhD on Computer Science (2015) from the UPC. From 2005 to 2008 he worked as research support engineer at the UPC, and joined BSC at the end of 2008. He has contributed in the design and development of different tools and programming models for complex distributed computing platforms. He has published over 30 research papers in conferences and journals and he has been involved in several National and European R\&D projects (FP6, FP7 and H2020). He is member of a program committee of several international conferences, reviewer of journal articles and he was member of the Spanish National Grid Initiative panel. His current research interests are focused on parallel programming models for heterogeneous parallel distributed computing environments and the interoperability between distributed computing platforms.
\endbio

\bio{pics/francesc}
Francesc Lordan obtained the Ph.D. in Computer Architecture from the Universitat Politecnica de Catalunya in 2018 after defending his Ph.D. thesis: "Programming Models for Mobile Environments". Since 2010, Francesc is part of the Workflows and Distributed Computing group of the Barcelona Supercomputing Center. His efforts have focused on the COMPSs programming model: a task-based model for developing parallel applications running on large distributed infrastructures such as clusters, supercomputers, grids and clouds. Francesc has published more than 20 articles in International conferences and journals, and he has been directly involved in the European projects mF2C, ASCETIC and OPTIMIS. He has also provided support to other collaborative projects such as Venus-C, EU-Brasil OpenBio, Transplant and the Human Brain Project. His research focuses on programming models that aim to ease the development of parallel applications by hiding the technical concerns of heterogeneous and distributed infrastructures.
\endbio

~\\
~\\
~\\
~\\
~\\
~\\

\bio{pics/rosa}
Rosa M. Badia  holds a PhD on Computer Science (1994) from the Technical University of Catalonia (UPC). She is the manager of the Workflows and Distributed Computing research group at the Barcelona Supercomputing Center (BSC). She is also a lecturer at the Technical University of Catalonia. Her current research interest are programming models for complex platforms (from edge, fog, to Clouds and large HPC systems) and its convergence with big data analytics and artificial intelligence. The group led by Dr. Badia has been developing StarSs programming model for more than 10 years, with a high success in adoption by application developers. Currently the group focuses its efforts in PyCOMPSs/COMPSs, an instance of the programming model for distributed computing, and its application to the development of large heterogeneous workflows that combine HPC, Big Data and Machine Learning. Dr Badia has published near 200 papers in international conferences and journals in the topics of her research. She has been very active in projects funded by the European Commission, participating in around 20 projects and in contracts with industry (Fujitsu, IBM and Intel). She has been actively contributing to the BDEC international initiative and is a member of HiPEAC Network of Excellence.
\endbio

\end{document}